\documentclass[a4paper,fleqn,usenatbib]{mnras}
\usepackage{newtxtext,newtxmath}

\usepackage[T1]{fontenc}
\usepackage{ae,aecompl}
\usepackage{lmodern}
\usepackage{graphicx}	
\usepackage{amsmath}	
\usepackage{amssymb}	
\usepackage{geometry}	

\newcommand{\bpass}{{\sc BPASS}}

\title[Habitability through Merger Trees]{Exploring the Cosmic Evolution of Habitability with Galaxy Merger Trees}
\author[E. R. Stanway et al.]{
E.~R.~Stanway,$^{1,2}$\thanks{E-mail: e.r.stanway@warwick.ac.uk}
M.~J.~Hoskin,$^{1,2}$
M.~A.~Lane,$^{1,3}$
G.~C.~Brown,$^{1}$
H.~J.~T.~Childs,$^{1}$\newauthor
\ S.~M.~L.~Greis$^{1}$
and A.~J.~Levan$^{1}$
\\
$^{1}$Physics Department, University of Warwick, Gibbet Hill Road, Coventry, CV4 7AL, UK\\
$^{2}$Centre for Exoplanets and Habitability, University of Warwick, Gibbet Hill Road, Coventry, CV4 7AL, UK\\
$^{3}$CANES program, Department of Physics, King's College London, London, WC2R 2LS, UK.
}

\date{Accepted 2017 December 13. Received 2017 December 1; in original form 2017 August 7}

\pubyear{2017}

\begin{document}
\label{firstpage}
\pagerange{\pageref{firstpage}--\pageref{lastpage}}
\maketitle

\begin{abstract}
We combine inferred galaxy properties from a semi-analytic galaxy evolution model incorporating dark matter halo merger trees with new estimates of supernova and gamma ray burst rates as a function of metallicity from stellar population synthesis models incorporating binary interactions. We use these to explore the stellar mass fraction of galaxies irradiated by energetic astrophysical transients and its evolution over cosmic time, and thus the fraction which is potentially habitable by life like our own. We find that 18 per cent of the stellar mass in the Universe is likely to have been irradiated within the last 260\,Myr, with GRBs dominating that fraction. We do not see a strong dependence of irradiated stellar mass fraction on stellar mass or richness of the galaxy environment. We consider a representative merger tree as a Local Group analogue, and find that there are galaxies at all masses which have retained a high habitable fraction ($>$40 per cent) over the last 6\,Gyr, but also that there are galaxies at all masses where the merger history and associated star formation have rendered galaxies effectively uninhabitable. This illustrates the need to consider detailed merger trees when evaluating the cosmic evolution of habitability.
  \end{abstract}

\begin{keywords}
astrobiology -- methods:numerical -- galaxies:evolution
\end{keywords}

\section{Introduction}\label{sec:intro}

The concept of extraterrestrial habitability, the possibility that there are regions beyond Earth's atmosphere where life like our own could survive, is of interest both to exoplanetary astronomers and to a far wider community. Understanding the potential for life elsewhere informs our perception both of our own place in the Universe and our relation to our own, eminently-habitable planet. However defining habitability is far from straightforward. The survival of life on Earth, its nature and evolution, has been shaped by the planet's geology and magnetic field, by its atmosphere and by the intensity and spectrum of irradiation from the Sun. However the progress of life on Earth, and elsewhere in the Universe, could also have been influenced by events outside the Solar System - an idea proposed as early as the 1950s \citep{krasovsky}. The explosions that mark the end of the life of massive stars - supernovae (SNe) and gamma-ray bursts (GRBs) - release hard radiation capable of affecting the biology of planets well outside their own solar systems. Effects from such events could include cosmic ray bombardment and the photodissociation and ionization of molecules in a planetary atmosphere \citep[e.g N$_2$ and ozone,][]{1995ApJ...444L..53T,1995PNAS...92..235E,1999NewA....4..419F}. Stripping of the atmosphere would increase the hard radiation reaching the planetary surface and this could cause cellular damage and disrupt DNA (or its analogues). Indeed, at least one mass extinction event in the history of life on Earth, in the late Ordovician period, has been attributed to the effects of a nearby GRB \citep{2004IJAsB...3...55M}, although this remains controversial. There is also strong evidence for irradiation by supernovae to be found on Earth, including identification of radioisotopes which indicate Earth was blanketted by cosmic rays as recently as 5\,Myr ago \citep{1999NewA....4..419F}. While rarer than stellar explosions, an episode of accretion onto the central supermassive black hole in a galaxy (creating an active galactic nucleus, or AGN), produces a hard radiation field sustained over a far longer period and so will also have biological implications which need to be reckoned into this complex picture \citep[see][and references therein]{1981Icar...46...94C,2011AsBio..11..343M}.

Thus analysis of potential habitability in the present day ($z=0$) Universe requires an extension beyond the conventional concept of a stellar `habitable zone', defined as the region within any given solar system in which liquid water could potentially exist on the surface of an Earth-like planet. The addition of constraints on where enough metals (elements heavier than helium) exist to form terrestrial planets, and on where those planets are likely to escape ionization damage from nearby astrophysical transient events, leads to the definition of a `Galactic Habitable Zone' \citep[e.g.]{2001Icar..151..307L,2001Icar..152..185G} identifying regions where not just the solar system but also the solar neighbourhood are hospitable to life like our own.

Studies of the the Galactic Habitable Zone of the Milky Way have developed from estimates of the distribution of metals and its effect \citep[e.g.][]{2004Sci...303...59L,2014PhRvL.113w1102P,2016ApJ...832...38G}, to analysis of cellular automata \citep{2012OLEB...42..347V} and detailed, spatially resolved hydrodynamic simulations of disk evolution and star formation. These have been used to study the shape and distribution of potentially habitable zones of the Milky Way and M33 \cite[e.g.][]{2016MNRAS.459.3512V,2017IJAsB..16...60F}. They suggest that the outskirts of the stellar disk are most likely to be habitable in today's massive spirals, a reassuring result given the Earth's location 8\,kpc from the centre of the Milky Way. These simulations generally
apply relatively simple analytic prescriptions for habitability with metallicity and local supernova rate in order to gain insight into the properties of massive disk galaxies. They provide little information on the habitability of galaxies in different mass ranges or environments.

This further aspect of habitability - on galactic rather than simply stellar scales of space and time - has been explored by a number of studies which vary in their findings.  An analysis by \citet{2002ApJ...566..723S} based on the local cosmic star formation rate and estimates for GRB and supernovae determined that biologically-significant irradiation events may occur here on Earth at the rate of hundreds per gigayear. While the vast majority of these are too minor to substantially impact complex life, the authors suggested that they contribute to the rate of evolution through genetic mutation.

\citet{2015ApJ...810...41L} used analytic formalisms for the GRB rate as a function of metallicity and redshift in the Milky Way to explore the frequency of more severe, life-destroying GRBs affecting Earth, finding $\sim1$ event in each 500\,Myr period, consistent with the interval since the Ordovician extinction event and other estimates of this rate \citep[see][]{2011AsBio..11..343M}. However they also evaluated the likely distribution of habitability at each galaxy's half-light radius (a scale length encompassing half the stellar light, and probable mass) based on the properties of galaxies in the Sloan Digital Sky Survey and an extrapolation of these to higher redshifts. They find that 50 per cent of galaxies at $z=1.5$ (a lookback time of 9\,Gyr) and 10 per cent at $z=3$ (11\,Gyr) remain habitable at the galaxy half light radius, based on analytic prescriptions for the evolution of star formation rate (SFR) per unit stellar mass in the galaxies, metallicity and the local relation between these quantites and typical galaxy scale length.

Similarly, \citet{2015ApJ...810L...2D} used analytic prescriptions for the fundamental mass-SFR-metallicity relationship observed in the local Universe to estimate the number of terrestrial planets and their rate of irradiation by supernovae in galaxies of different masses. Based on volume-averaged estimates in the local Universe, they suggest that giant elliptical galaxies (which have high metallicities and low star formation rates) may be the most secure cradles of life at $z=0$. An extension to this work \citep{2016arXiv160609224D} incorporated prescriptions for the volume averaged SFR density history to examine how their conclusions changed over cosmic time. They found that supernovae dominate the threat to habitability at all times, but made the simplifying assumption that all transient events irradiate the same basic volume (with a reduction factor to account for relativistic beaming in the case of GRBs).   Given this (volume-averaged) prescription, the mean habitability in the Universe has increased by a factor of 2.5-20 over the last 4\,Gyr.

Another analytic model presented by \citet{2016A&A...592A..96G} also constrained habitability through joint constraints on terrestrial planet formation and current rates of astrophysical transients, combined with estimates of the volume-averaged SFR density history. However their conclusion differed significantly from that of \citet{2016arXiv160609224D}, suggesting that the mean habitability of galaxies has not changed significantly over the last 8\,Gyr. In both cases, the rate of SNe and GRBs was determined by scaling the star formation rate by a fixed fraction, identifying the stars expected to be transient progenitors in simple stellar evolution models at Solar metallicity.

While this work provides a valuable insight into the volume-averaged evolution of cosmic habitability through analytic means, and the uncertainties arising from assumptions, it is difficult to constrain the distribution of galaxy habitability or its detailed dependence on formation history through such methods. Here we take an intermediate approach between the detailed high resolution studies of the Milky Way galactic habitable zone and the semi-analytic volume-averaged cosmic evolution models discussed above. In this paper we exploit the power and resolution of a large dark matter halo merger simulation, the Millenium Simulation \citep{2005Natur.435..629S}, and the detailed star formation and merger histories predicted in an associated semi-analytic galaxy formation model  \citep{2007MNRAS.375....2D}. We combine this with a new detailed stellar population synthesis model incorporating the effects of stellar binary evolution with metallicity \citep{E17},  to estimate the energetic transient event rates of a large sample of individual simulated galaxies. From these we determine the fraction of each galaxy likely to be strongly irradiated in a given epoch and hence the potentially habitable fraction as a function of stellar mass and galaxy environment across cosmic time.

As will have become clear, habitability is a flexible term with applications in contexts from the largest cosmic scales down to the occupancy of individual buildings on Earth. Here we consider a simple definition through negation: we define a region as potentially habitable if it has {\it not} been irradiated by one or more nearby energetic events such as a supernova (SN), gamma-ray burst (GRB) or active galactic nucleus (AGN). For each galaxy in a semi-analytic model of galaxy formation, we define the mass fraction of the galaxy which satisfies this criterion in a given time interval, and which is also likely to host terrestrial planets, as $H_\mathrm{rad}$, as discussed in section \ref{sec:habcriteria} below.

 In section \ref{sec:method} we present our methodology and introduce the models we employ. In section \ref{sec:volavg} we consider the distribution of $H_\mathrm{rad}$ in a representative cosmic volume. together with its dependence on galaxy parameters. In section \ref{sec:casestudy} we focus as a case study on a galaxy merger tree that identifies a potential Local Group analogue at the current day. In section \ref{sec:discussion} we discuss our results and their implications in the wider context of studies of galactic habitability, before summarising our conclusions in section \ref{sec:conc}.

For consistency with the semi-analytic models on which this analysis is built, we use the standard cosmology of the Millenium Simulation: [$\Omega_m,\Omega_b,\Omega_\Lambda,h$]=[0.25,0.045,0.75,0.73] 


\section{Methodology}\label{sec:method}

 \subsection{The Millenium Simulation and Semi-Analytic Models}\label{sec:mill}

 The Millenium Simulation \citep{2005Natur.435..629S} was a cosmological N-body simulation designed to trace the evolution and merger history of dark matter halos from $z\sim100$ to the present day. The full simulation comprised 2160$^3$ particles in a periodic 500\,Mpc/$h$ box, while the smaller `millimil' simulation comprised 270$^3$ particles in a 62.5\,Mpc/$h$ box. Each particle represents a finite mass of dark (non-baryonic, gravitationally-interacting) matter. By tracing these dark matter particles through time and space, and applying a variety of group-finding algorithms, individual halos of gravitationally bound particles and subhalos within those structures can be identified. This allows the evolution of the massive dark matter halos, and hence of large scale structure in the Universe, to be studied in detail.

 The catalog of halos and associated subhalos in the Millenium Simulation, together with their properties, were recorded at 63 time intervals, referred to as {\it snaps} (where snap 0 lies at the highest redshift and snap 63 represents the current time at $z=0$). These are spaced to follow the evolution of haloes, and so are separated by intervals as short as 10\,Myr at early times when the Universe was rapidly evolving, but are typically of order 200-300\,Myr at $z<3$. The evolution and association of halos between time steps was recorded through the use of merger trees. In this formalism, the descendants and progenitors of any given halo are identified in adjacent snaps and their identities recorded. This allows all the halos which will eventually contribute to a single massive galaxy at the current time to be identified, and their mass growth and interactions monitored. 

 Baryonic matter, from which stars and planets form, is gravitationally attracted to dark matter, and so gathers in the potential well of dark matter halos. In these regions, the diffuse intergalactic medium becomes concentrated and gradually cools, allowing the formation of stars and galaxies. The properties of these galaxies are shaped first by the dark matter gravitational potential, and secondly by feedback processes in which energy is returned to the environment such as supernovae and stellar winds. Studies of galaxies and galaxy populations over cosmic time allow the determination of analytic relations between dark matter halo properties and baryonic matter properties, as well as empirical corrections for non-analytic relations such as the feedback energy released by a given supernova.  These empirical and analytic rules are combined to form what are known as 'semi-analytic' models which relate the properties of a galaxy both to its associated dark matter halo and to the properties of the galaxy and its environment in earlier and later timesteps. The resulting model galaxies will accrete material from their surroundings, return material to the intergalactic medium, form stars, increase in metallicity, acquire central supermassive black holes and merge with one another over time.  While galaxies in dark matter halos of comparable mass may begin with similar baryonic properties, these will diverge as the evolution takes into account individual events such as halo mergers and episodes of accretion onto a central massive black hole.

 As such a semi-analytic merger tree formalism allows the diversity of galaxy properties at a given mass, metallicity or cosmic time to be explored. Any given merger tree results ultimately in the formation of one giant galaxy, or one galaxy cluster in the case of large halos, at the current time, while using all the merger trees in a representative cosmic volume explores different pathways which may lead to the formation of superficially similar galaxies at the current time. The versatility of this formalism, and its potential for in depth understanding of the effects of galaxy evolution on the stars in a given galaxy, have not hitherto been applied to a study of habitability. This is largely due to the limited mass resolution of such models, which do not allow detailed study of habitable zones within a single galaxy, but are rather designed to explore the global properties of a galaxy population (although this is changing as both the time and mass resolution of such models improve).

 The Millenium Simulation dark matter halo merger trees have been used as a framework on which to base a number of different semi-analytic galaxy evolution models. Here we use the work of \citet{2007MNRAS.375....2D} which uses the L-Galaxies evolution model \citep{2006MNRAS.365...11C,2006MNRAS.366..499D} to trace the properties and results of star formation, and incorporates a variety of feedback processes including radio mode feedback. The \citet{2007MNRAS.375....2D} galaxy catalog was optimised to trace the formation of brightest cluster galaxies, and considers their progenitors and merger histories. Within the semi-analytic prescription, each halo-halo merger is accompanied by a collisional starburst in which gravitational interactions between galaxies leads to the compression of gas through shocks and ram pressure, resulting in an upsurge in the instantaneous rate of formation for stars. Since the most massive stars are short lived, this will be immediately followed by a similar surge in the supernova rate, which returns both heavy elements and dust to the interstellar medium. Following a starburst, the star formation rate, chemical evolution, gas fraction, dust content and stellar mass evolution of individual galaxies is adjusted and tracked accordingly. The semi-analytic model prescriptions are tuned to match the joint distribution of galaxy luminosity, colour and morphology in the local Universe. They adopt the same cosmology as the Millenium Simulation as described above.
 
 For the purposes of this pilot study we make use of the `millimil' database\footnote{Accessed at http://gavo.mpa-garching.mpg.de/Millennium/}, which presents a subset of the full simulation.  While this models a smaller volume than the full simulation, and so does not contain the most massive cluster haloes,  the millimil subset is sufficiently large that taking the mean of model galaxy properties within a given timestep allows users to average over cosmic variance (i.e. the probability of studying an over- or under-dense region of the cosmic web) and reproduces a smoothly evolving cosmic star formation rate density function.

\begin{figure}
  \includegraphics[width=\columnwidth]{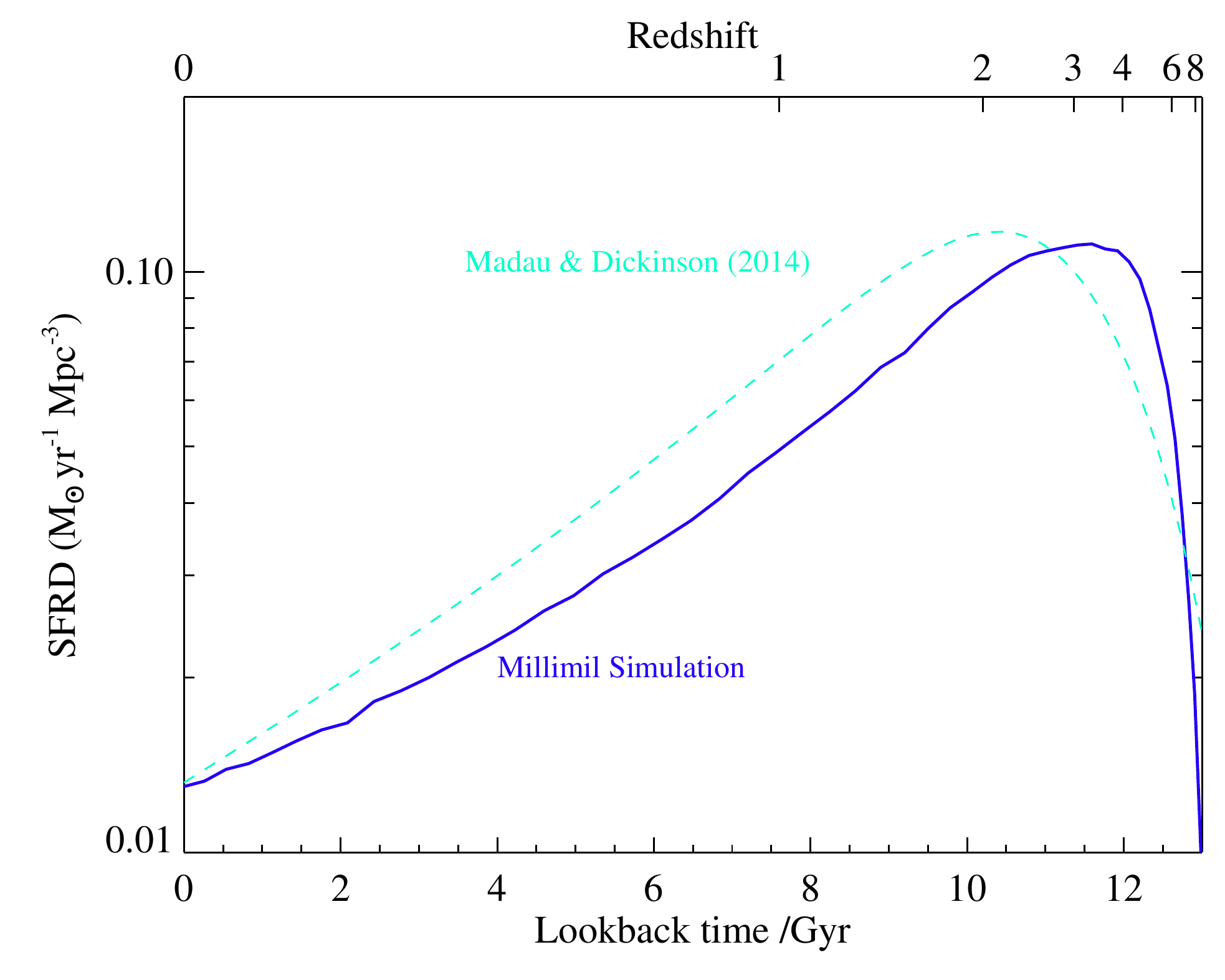} 
\caption{ The star formation rate density (SFRD) history of galaxies in the millimill simulation box. We compare this against the recent analytic form for the SFRD function proposed by  \citet{2014ARA&A..52..415M}.}\label{fig:md14sfh}
 \end{figure}

\begin{figure*}
  \includegraphics[width=1.9\columnwidth]{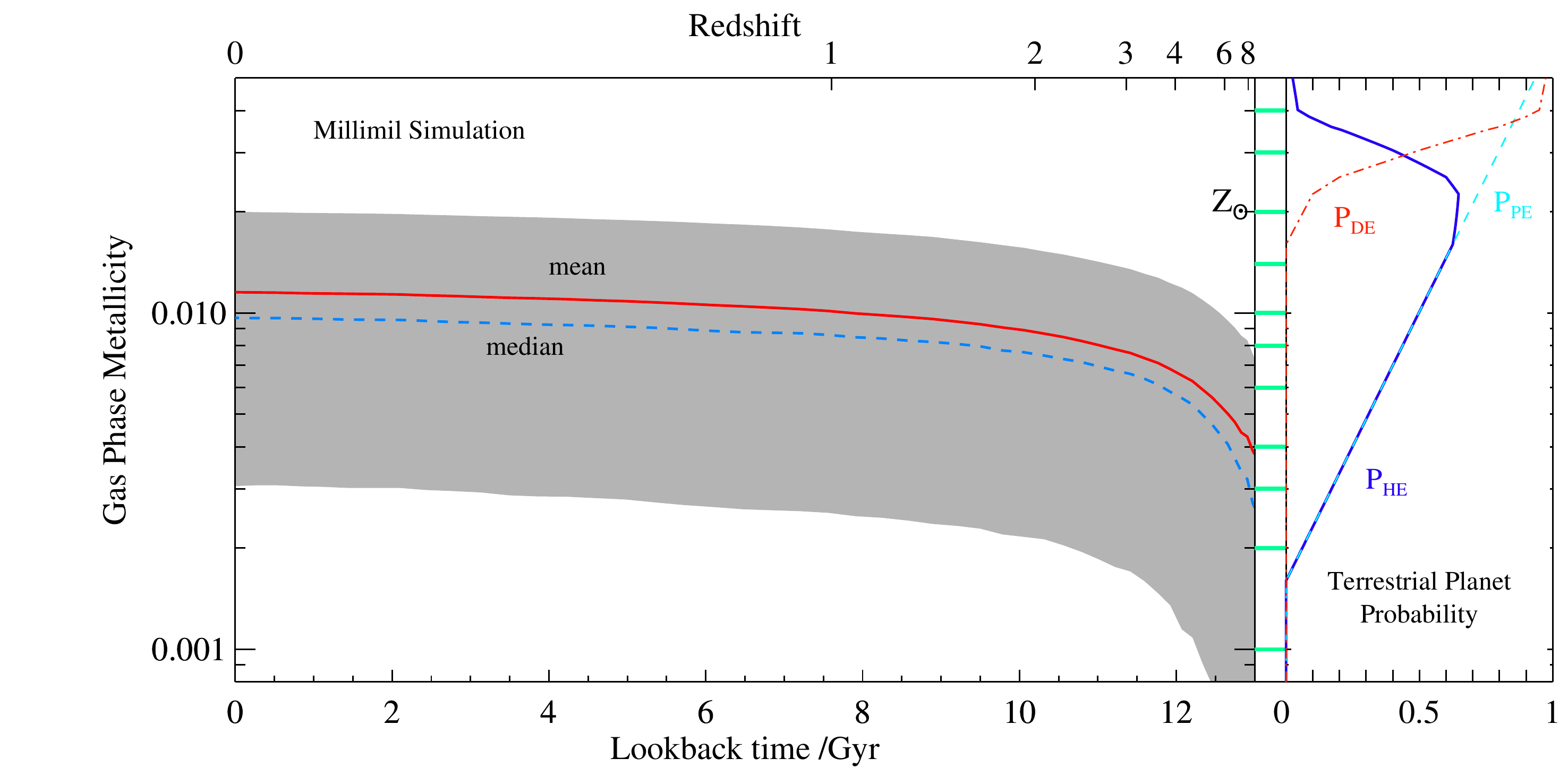}
  \caption{The evolution of cool gas phase metallicity in millimil galaxies. We show the mean and median metallicity at each snap. The shaded region indicates the $\pm1\,\sigma$ spread in the galaxy metallicity distribution. Horizontal bars in the central panel indicates the metallicities sampled by the \bpass\ transient rate estimates (see section \ref{sec:transient}). The left-hand panel indicates the probability of terrestrial planet production (P$_\mathrm{PE}$), destruction (P$_\mathrm{DE}$) and hosting (P$_\mathrm{HE}$), as estimated by Lineweaver (2001).}\label{fig:zevol}
 \end{figure*}

 As figure \ref{fig:md14sfh} demonstrates, the form of the cosmic star formation density history in the millimil galaxy database is broadly similar to the analytic form derived by \citet{2014ARA&A..52..415M} based on an extensive review of observational evidence, but is slightly narrower. It peaks at $z\sim3$ (c.f. $z\sim2$ for \citet{2014ARA&A..52..415M}) but at a comparable star formation density. The effect of this will be to lead to us estimating a lower transient rate and so a smaller irradiated fraction than might be derived from the  \citet{2014ARA&A..52..415M} star formation history, so our values can be treated as lower limits. The mean gas phase metallicity in galaxies also evolves smoothly in the millimill sample, as indicated by figure \ref{fig:zevol}, but the metallicity at any given lookback time shows a large spread in values. The joint effect of the galaxy mass distribution and  mass-metallicity relation \citep{2004ApJ...613..898T} is to produce more galaxies below the mean metallicity than above it at any given time, biasing the median galaxy to a lower metallicity than the mean galaxy.

\begin{figure}
  \includegraphics[width=\columnwidth]{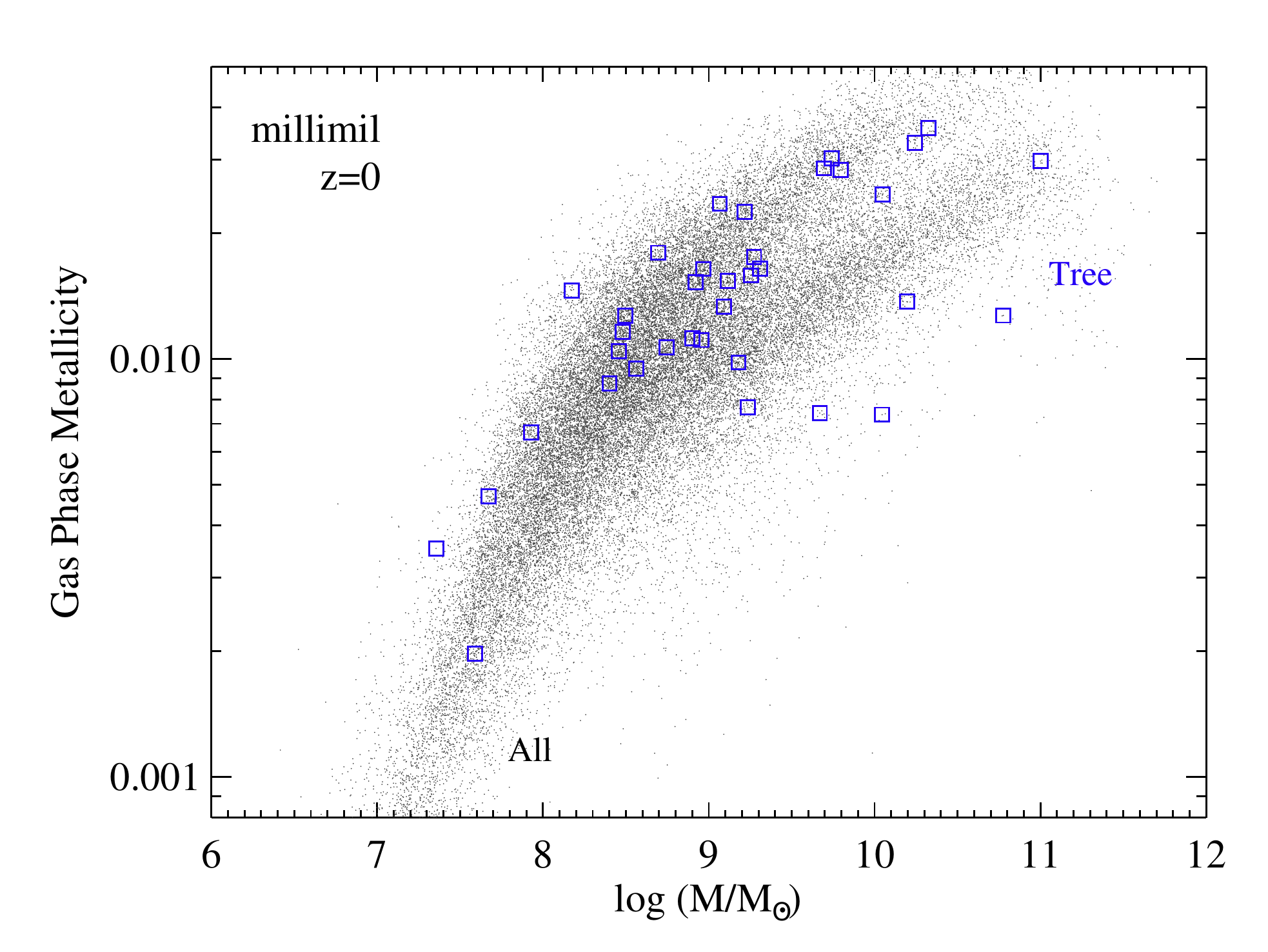} 
  \caption{The distribution of stellar masses and metallicities in the millimil survey database at $z=0$ (snapnum 63). We show the distribution of all the galaxies in the 60\,Mpc/$h$ cube as grey ponts, and also the 36 galaxies occupying the Local Group analogue halo merger tree discussed in section \ref{sec:casestudy} as blue squares.}\label{fig:massdist}
 \end{figure}

In figure \ref{fig:massdist} we show the distribution of $z=0$ (snap 63) stellar masses and metallicities in the simulation galaxy catalog. The millimill subset contains galaxies as small as $10^6$\,M$_\odot$ and extending up to a present day stellar mass of log($M$/M$_\odot$)$\sim$11.5 (comparable to the Milky Way) and also contains merger trees over a broad range of halo masses, allowing studies of Local Group analogues (see section \ref{sec:casestudy}).

 \subsection{Transient Event Rates}\label{sec:transient}

 We combine this information with version 2.1 of the Binary Population and Spectral Synthesis (BPASS) models to predict the rate of potentially-destructive astrophysical transients.

 \bpass\ models \citep{2009MNRAS.400.1019E,2012MNRAS.419..479E,2016MNRAS.456..485S,E17}\footnote{Available at http://bpass.auckland.ac.nz} predict the evolution of a simple stellar population of co-eval stars, distributed according to a broken power law initial mass function such that $N(M) \propto M^{-1.3}$ in the mass range $0.1<M<0.5$\,M$_\odot$ and $N(M) \propto M^{-2.35}$ at  $0.5<M<300$\,M$_\odot$. They incorporate the effects of binary mass transfer on stellar evolution, allowing an estimate of the supernova rates arising as a function of population age and metallicity. These include preliminary estimates of both Type Ia supernova and GRB rates.
 
\begin{figure}
  \includegraphics[width=\columnwidth]{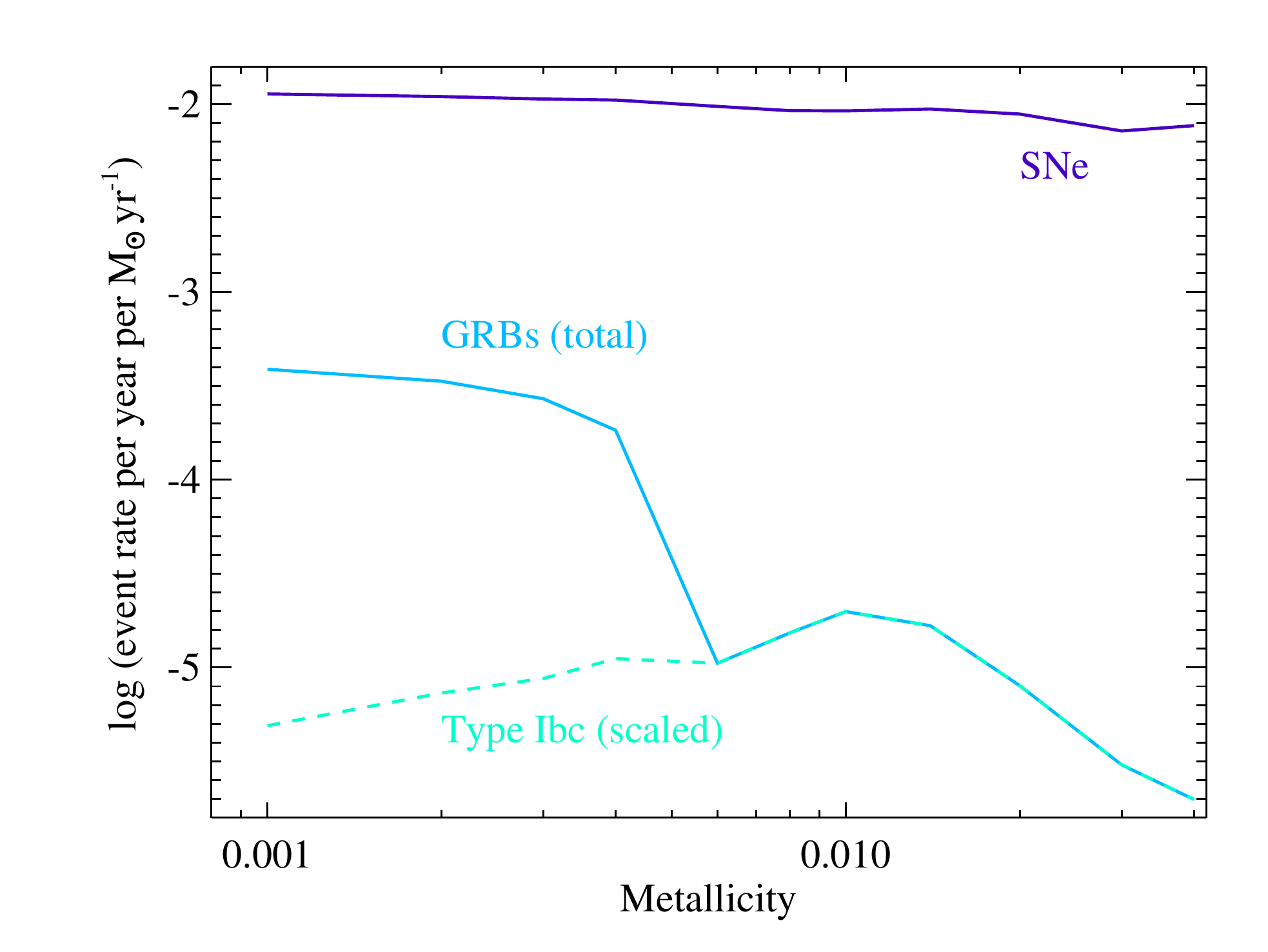} 
  \caption{The transient event rates predicted by \bpass\ for a population forming stars at a constant rate of 1\,M$_\odot$\,yr$^{-1}$. The supernova rate includes predictions for Type Ia,b,c and Type II events. The total GRB rate shown incorporates \bpass\ predictions based on rotationally mixed chemically homogenous evolution (which only occurs at low metallicity) together with a scaled fraction of the type Ibc events as discussed in the text.}\label{fig:transient}
 \end{figure}

 The snap intervals of the Millenium Simulation correspond to a typical elapsed time of $\sim$200-300\,Myr between steps (decreasing to shorter intervals at early times). We treat the star formation rate as constant in each interval and estimate the resultant event rate from a combination of simple stellar populations. In each case we select rates at the closest metallicity available in the model set to that determined by \citet{2007MNRAS.375....2D} for the cool gas in the galaxy. Supernova rate is mildly metallicity dependent, as figure \ref{fig:transient} demonstrates. We select the gas phase metallicity as this is more likely to be representative of the young stellar population generating supernovae than the bulk stellar metallicity.

 \bpass\ GRB event rates are based on an assumption of rotational mixing and subsequent chemically homogenous evolution that only occurs at $Z<0.2$\,Z$_\odot$. The metallicity dependence of GRB occurence is still widely debated \citep[see e.g.][]{2009ApJ...702..377K}, but it is clear that some events do occur at near-Solar metallicity which suggests that a progenitor population exists that is not fully accounted for in the model rates. We adjust for this by setting the GRB rate at metallicities below $Z=0.014$ (0.7\,Z$_\odot$) to 1 per cent of the Type 1bc event rate \citep{2012MNRAS.423.3049G,2007A&A...471..585B}. Above this metallicity we taper off the type Ibc fraction generating GRBs as ($Z$/0.014)$^{-2}$ to reflect the apparent metallicity bias in the GRB host population. At the lowest metallicities we boost this rate by the output of the rotationally mixed channel mentioned above. We do not consider the effects of short GRBs in this initial study since these are significantly less luminous than core-collapse-driven long GRBs.

 We note that \bpass\ models suggest that the SNe and GRBs arising from a population can continue for up to 1\,Gyr after the truncation of star formation, but also that the rate drops sharply by $2-3$ orders of magnitude at stellar population ages above $\sim100$\,Myr (Eldridge et al 2017). As a result, this provides a small perturbation on any ongoing star formation and we do not track the consequences of star formation in later snaps.

 \subsection{Habitability Criteria}\label{sec:habcriteria}

The \citet{2007MNRAS.375....2D} models estimated the stellar mass, star formation rate, gas-phase metal content, X-ray luminosity and virial size of individual galaxies at each timestep. To these we have added event rates of supernovae and GRBs. We use this information to constrain the galactic habitability as follows:

 \subsubsection{Metallicity Weighting}\label{sec:habZ}
 
 A key parameter in determining the potential habitability of a stellar system is its capacity to host terrestrial planets.  Previous work on the Galactic Habitable Zone has considered two key metallicity-dependant constraints:  the probability that a solar system at a given metallicity is able to form terrestrial planets, and the probability that a massive planet, migrating inwards through the exosolar system to become a hot Jupiter would eject any terrestrial planets that had formed. The first of these provides a lower metallicity cut-off (since rocky planets cannot form in a system lacking heavy elements), and the second represents a high metallicity truncation (since hot Jupiters appear to be more common in metal-rich systems).

 For consistency with previous work, we adopt the probability distributions proposed by \citet{2001Icar..151..307L}. In this prescription, the probability of producing Earth-like planets ($P_\mathrm{PE}$)  is proportional to metallicity $Z$ above a threshold of 0.1$Z_\odot$, rising to unity at log($Z$/$Z_\odot$)=0.6 where, in this case, $Z_\odot=0.016$. The probability of destroying Earth-like planets ($P_\mathrm{DE}$) rises sharply from 20\% to 90\% in the range  log($Z$/$Z_\odot$)=0.2 to 0.4. The probability of hosting Earth-like planets, $P_\mathrm{HE}$ is thus given by:
 \begin{equation}
   P_\mathrm{HE} = P_\mathrm{PE}\, \left(1-P_\mathrm{DE}\right).
   \end{equation}

 The resulting probability distributions are shown in the right-hand panel of figure \ref{fig:zevol}, and lead to a strong bias towards Solar and super-Solar metallicity systems as potential terrestrial planet hosts, while substantial tails of potentially habitable terrestrial planet systems remain down to low metallicities. Given the flat evolution in the mean gas phase metallicity of the Universe over cosmic time, this probability distribution has relatively little effect on the evolution of habitabilities at any lookback time $<10$\,Gyr, but instead acts primarily as a scaling factor, truncating the maximum habitable fraction at any redshift at $\sim50$\%. It will, however, tend to suppress the habitable fraction in low mass galaxies (log(M/M$_\odot$)$<$8.5), since these have lower typical metallicities as shown in figure \ref{fig:massdist}.

 We note that since \citet{2001Icar..151..307L} defined this probability distribution, the number of exoplanetary systems known has grown exponentially \citep[see e.g.][for a recent review]{2015ARA&A..53..409W}. Nonetheless, the metallicity dependance of planet hosting remains a complex subject, with relatively poor observational constraints, particularly at the low end of the planetary and stellar mass functions where observational selection effects are significant. Nonetheless the distribution above remains broadly consistent with current observations. \citet{2015AJ....149...14W} examined the distribution of planets selected by the {\it Kepler} telescope. They lacked the sample size to determine the detailed metallicity distribution of the planet hosts, but found that the median metallicity for terrestrial planet hosts is log($Z$/$Z_\odot$)=0.04 (with $Z_\odot=0.018$), which matches the peak of the \citet{2001Icar..151..307L} distribution in $P_\mathrm{HE}(Z)$.

 By contrast, other studies such as that of \citet{2016AJ....152....4A} and \citet{2012Natur.486..375B} have found no statistically significant difference in metallicity between stars hosting terrestrial planet candidates and the whole {\it Kepler} sample, even when only planets in their habitable zones are considered. \citet{2017arXiv171104878H} also suggest that, while M-dwarf stars hosting massive planets are biased towards high metallicities, no such bias is seen for the hosts of low mass planets. 

At the lowest metallicities, \citet{2016ApJ...833..214Z} have studied the history of metallicity enhancement in the Universe and its effect on terrestial planet formation.
 They apply a gradual iron enrichment cutoff between 10 and 1 per cent of Solar metallicity, extending well below the \citet{2001Icar..151..307L} cutoff we apply, although, as figure \ref{fig:zevol} demonstrates, the vast bulk of galaxies have a mean gas phase metallicity above 10 per cent Solar for the entire lookback interval being considered.

 It is likely that future exoplanet telescope missions such as {\it TESS} and {\it PLATO}, together with extensive ground-based follow-up, will be required to clarify this issue.  However, as noted above, given the relatively slow evolution of mean metallicities in cosmic time, we expect alternate prescriptions for planet hosting to result in a systematic rescaling of our habitability fractions for all but the lowest mass galaxies.

 \subsubsection{Irradiation}\label{sec:habrad}
 To evaluate the effect of irradiation, we model each galaxy as a spherical system, with the density of stars varying as $\rho_\ast(r) \propto \exp{(-r/R_\mathrm{eff})}$, where $r$ is measured from the centre of each galaxy and $R_\mathrm{eff}$ is estimated from the virial radius as $R_\mathrm{eff}=0.015\,R_\mathrm{vir}$ \citep[where $R_\mathrm{vir}$ is provided by the simulation and the scaling factor is derived from the local galaxy population by][]{2013ApJ...764L..31K}. We draw locations for transient events randomly from this stellar distribution to determine the distribution of local stellar densities in the environs of supernova and GRB transients. We assume (and have tested the assumption) that any galaxy with 1000 or more events fully samples this distribution in $r$/$R_\mathrm{eff}$.

 Each supernova is assumed to irradiate a spherical volume with a radius $r_\mathrm{SN}$. The fraction of the galaxy's stellar mass irradiated by each event is then given by the product of this volume with the stellar density distribution as described above, scaled by the stellar mass and effective radius of the host galaxy. We do not consider the effects of a luminosity distribution in supernovae (see section \ref{sec:discussion}) but treat each as irradiating the same volume. The effect of multiple supernovae in a galaxy in a snap interval is cumulative, with each removing an appropriate fraction of the remaining, non-irradiated stellar mass. The total irradiated fraction for each galaxy is $f_\mathrm{SN}$. We adopt $r_\mathrm{SN}=8$\,pc \citep{1995ApJ...444L..53T,2003ApJ...585.1169G}. 

 We calculate the fraction irradiated by GRBs, $f_\mathrm{GRB}$, using a near-identical method, drawing on the same underlying stellar density distribution. However GRB emission is collimated into a narrow jet and only stars along the jet-axis are likely to be strongly irradiated. The on-axis distance over which the GRB will have a severe impact on a planetary biosphere depends, of course, on the details of atmosphere, planetary conditions and exobiology. A useful criterion employed by several authors is the distance at which GRB irradiation is likely to deplete the protective ozone layer in an Earth-like atmosphere. As \citet{2005ApJ...622L.153T,2005ApJ...634..509T} discuss, serious biological consequences which, on Earth, would extend to destroying plankton organisms that support the food chain and triggering glaciation, are very likely at irradiation distances $d\sim1-2$\,kpc. To err on the conservative side, we adopt an irradiated radius $r_\mathrm{GRB}=1$\,kpc but reduce the spherical volume to only consider the solid angle corresponding to a jet opening angle of 10\,degrees \citep[see][]{2016ApJ...818...18G}. As a result, GRBs irradiate a volume $\sim$1900 times larger than a supernova in the same galaxy.
 
Finally we consider the possible influence of AGN activity. The \citet{2007MNRAS.375....2D} catalog records possible AGN accretion in the form of a predicted X-ray luminosity. We use this to set a flag assigning the AGN as `on' in a snap if $\log(L_X / \mathrm{ergs\,s}^{-1})>36$ and `off' otherwise. If the AGN is on, we take a conservative approach and assume stars in the galactic core (distributed in $r$ as above) are irradiated out to a radius $r_\mathrm{AGN}=100$\,pc, giving an irradiated fraction $f_\mathrm{AGN}$. 

We discuss uncertainties on the parameters selected here in section \ref{sec:discussion} and in Appendix \ref{appendix}.

\subsubsection{Defining Habitability}\label{sec:habdef}

We designate the fraction of a galaxy that remains potentially habitable accounting for irradiation, $H_\mathrm{rad}$, as the mass fraction that both can host a terrestrial planet and is not irradiated by an energetic event in the {\it current} snap interval. This quantity is defined as:
\begin{equation}
  H_\mathrm{rad} = P_\mathrm{HE}\, f_\mathrm{n} =P_\mathrm{HE} \,\left(1-f_\mathrm{AGN}\right) \, \left(1-f_\mathrm{SN}\right) \, \left(1-f_\mathrm{GRB}\right)\,.
\end{equation}

As discussed above, this quantity is calculated for irradiation within the time step in question (typically $\sim$200\,Myrs below $z\sim3$) so irradiation in a previous timestep is no bar to habitability for rapidly-arising life. It is also designed to avoid double-counting mass fractions irradiated more than once in the same timestep, instead it identifies the fraction {\it not} irradiated in the current timestep, and its complement.

We do not typically follow the effect of star formation into later timesteps, since the decline in transient rate is rapid after the cessation of star formation. An exception occurs when we apply a {\it habitability timescale}, $t_\mathrm{h}$ which we provisionally set at 1\,Gyr. In this instance we record the fraction of stars that have been irradiated within $t_\mathrm{h}$ by following galaxy merger trees, such that the revised potential habitability at a given timestep is
\begin{equation}
    H_{\mathrm{rad,t_h}} = f_\mathrm{n} \left(P_\mathrm{HE} \psi \Delta t + \Sigma_i H_{\mathrm{rad},i} M_{\ast,i}\right)  M_\ast^{-1}\,
  \end{equation}
where the summation is over progenitor galaxies at the previous snap intervals and  $P_\mathrm{HE}$, $M_\ast$ and $\psi$ are the terrestial planet host fraction, stellar mass and star formation rate in the current interval, which has a duration $\Delta t$. This algorithm is applied iteratively forward in time to each timestep within $t_\mathrm{h}$. We note that a few galaxies actually lose mass between timesteps, presumably due to some form of stripping. In these cases we do not add additional stellar mass during the summation.

\section{Volume Averaged Habitability}\label{sec:volavg}

\begin{figure}
  \includegraphics[width=\columnwidth]{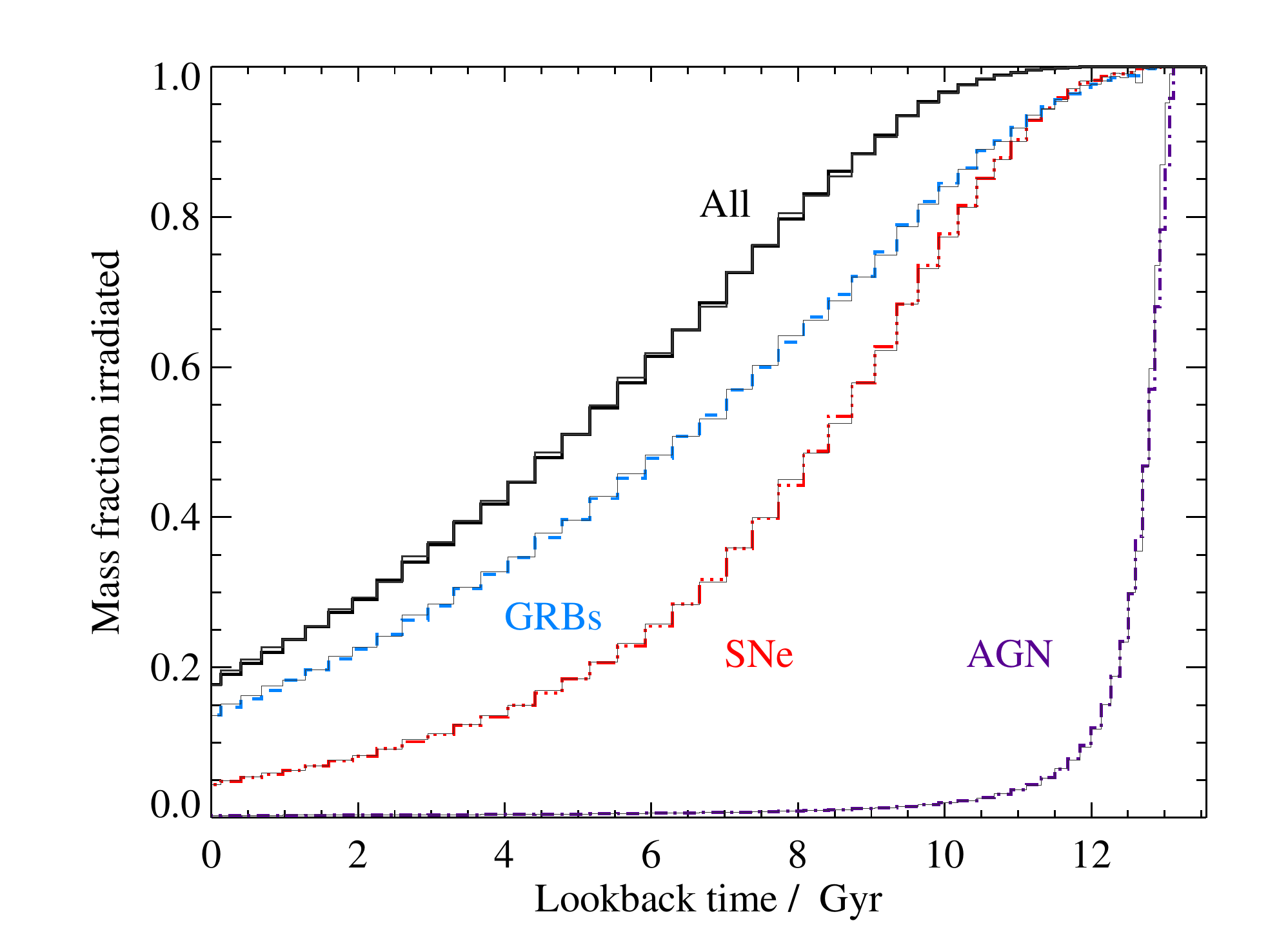} 
  \caption{The fraction of the stellar mass in each simulation time interval within our modelled volume that is irradiated by at least one supernova (red, dot-dot-dashed line), at least one GRB (blue, dashed), or at least one AGN (purple, dot-dash) or at least once by any event (black, thick solid line) within the snap interval in question. Each step in the curves indicates a different snap, these are typically of length $\sim$200\,Myr. Thin solid lines indicate the equivalent fraction of stellar mass capable of hosting terestrial planets that is irradiated (i.e. applying the metallicity weighting in figure \ref{fig:zevol}).}\label{fig:irrtime}
 \end{figure}

We consider first the volume-averaged habitability as a function of time by considering the bulk properties of the entire millimil-derived dataset. In figure \ref{fig:irrtime} we illustrated the fraction of the total stellar mass irradiated by one or more SNe, GRBs and AGN in each snap interval (i.e. $f_\mathrm{n}$). We find that 4.4 per cent of stellar mass in the simulation volume has been irradiated by supernovae in the current time interval (snap 63, i.e. the last $\sim$260\,Myr), while 14 per cent has been irradiated by GRBs, and 0.3 per cent by AGN. These fractions rise sharply with lookback time. Allowing for overlapping regions of influence, 18 per cent of the stellar mass in galaxies is irradiated. These distributions are unchanged when the metallicity-weighting $P_\mathrm{HE}$ is taken into account to find the fraction of the {\it terrestrial planet-hosting} mass irradiated (shown as thin black lines on the figure). This reflects the fact that the metallicity distribution of galaxies is effectively constant over the bulk of cosmic time considered here, so the planet-hosting mass fraction is near constant. As a result, the volume-averaged fractions of  total stellar mass irradiated and  planet-hosting mass irradiated vary together. While the  habitable fractions of individual galaxies will depend on their own mass and metallicity evolution, the volume-averaged quantities are sensitive primarily to the volume-averaged mean metallicity and star formation rate.

The tendency for both SNe and GRBs to irradiate a lower mass fraction of stars in the local Universe compared to at high lookback times reflects several aspects of cosmic evolution: the volume-averaged star formation rate is lower and the metallicity higher (as shown in figures \ref{fig:md14sfh} and \ref{fig:zevol}), both of which act to reduce the transient event rates. However these trends are reinforced by the effects of heirarchical galaxy mergers; more stars are concentrated in the most massive galaxies which tend to be more extreme in both star formation rate and metallicity, so the stellar density in irradiated volumes at late times will itself be lower. Similarly, the effects of galaxy evolution can be seen at early times: while the volume-averaged star formation rate drops dramatically at lookback times $\gtrsim 11$\,Gyr, those stellar explosions that do occur take place in much smaller galaxies. Since AGN activity is modelled as irradiating a constant volume, this will also encompass a larger fraction of each galaxy at early times if the AGN is active. As a result a far larger fraction of the stellar mass in each galaxy is irradiated by each event at early times.

We note that although GRBs are far less frequent than supernovae at any given star formation rate (see figure \ref{fig:transient}), they take over as the dominant source of irradiation fairly early in the history of the Universe (i.e. at lookback times $<11$\,Gyr). Again this is primarily an effect of galaxy size evolution. The volume irradiated by a GRB is significantly larger than that irradiated by a supernova. As the average galaxy increases in physical extent, the irradiated stellar mass fraction resulting from each GRB decreases at a slower rate than that for supernovae as a result. For most of the history of the Universe, the contribution of GRBs and SNe to irradiation of their host galaxies is comparable, differing by a factor of a few.

\begin{figure*}
  \includegraphics[width=\columnwidth]{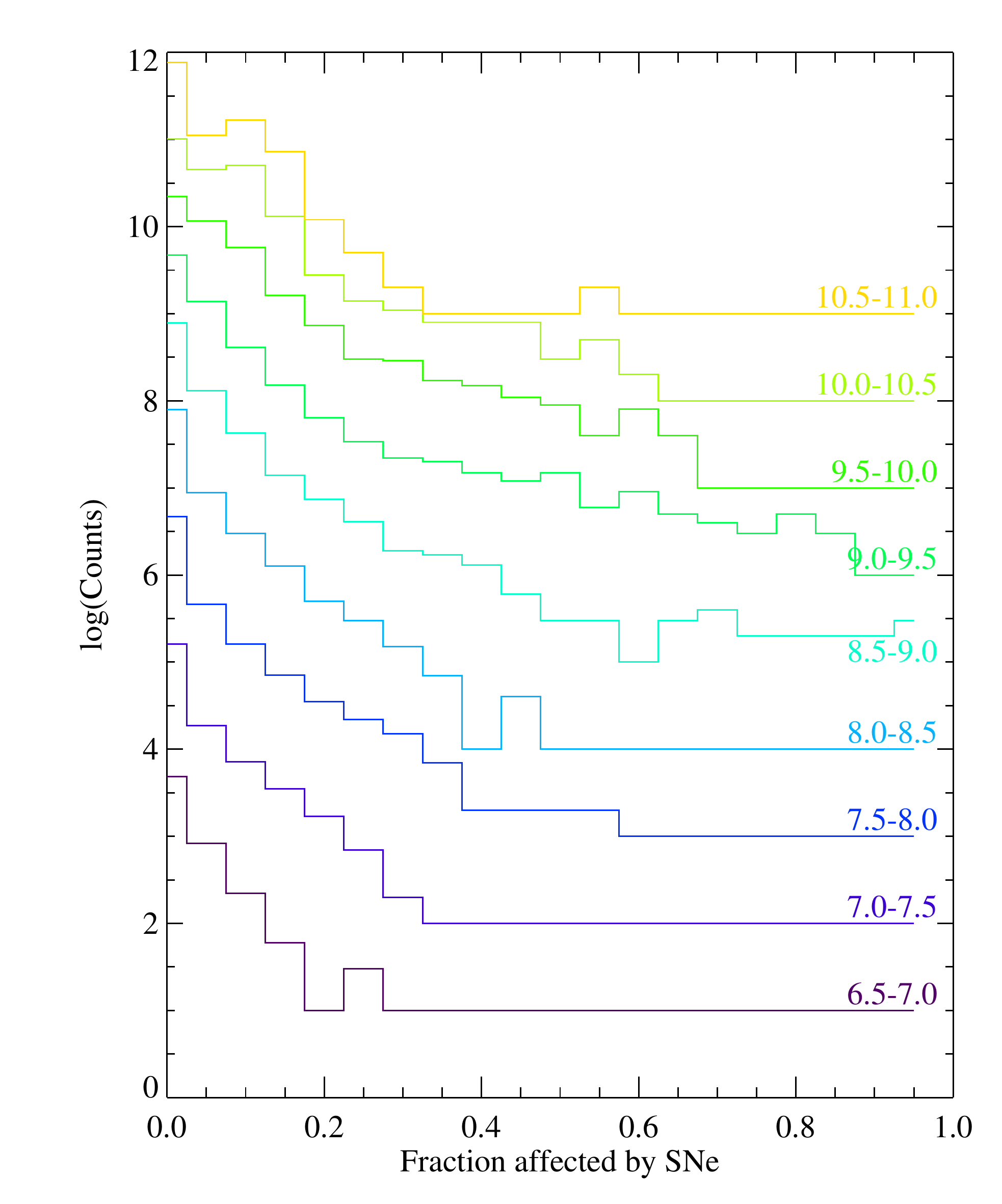} 
  \includegraphics[width=\columnwidth]{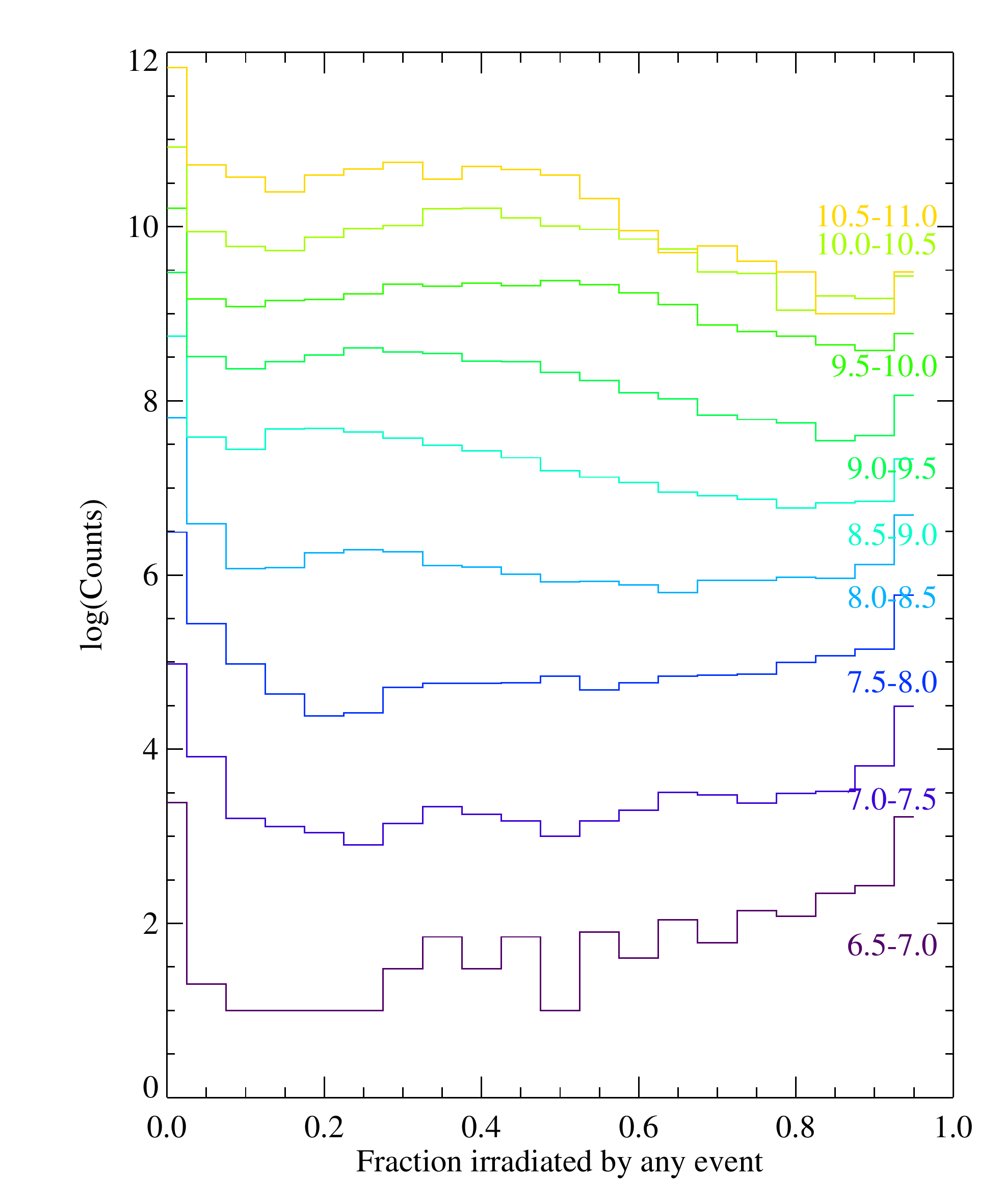} 
  \caption{The number of galaxies at the current time with a given irradiated mass fraction, as a function of galaxy stellar mass. The left-hand panel considers only irradiation by supernovae, while the right hand panel considers SNe, GRBs and AGN. Successive mass bins are offset by 1 dex.
  }\label{fig:irrmass}
 \end{figure*}

 Given that galaxy size has a strong affect on the fraction of the stellar mass irradiated by each event, we also consider the dependence of irradiated mass fraction on galaxy stellar mass at the current time (i.e. irradiated by events at $z=0$, snap 63). When the effects of supernovae alone are considered (figure \ref{fig:irrmass}, left), galaxies in any given mass bin show a range of irradiated fractions, but with the bulk of the population biased towards low irradiated fractions. The strength of this bias decreases with increasing stellar mass as the star formation in the high mass population becomes more stochastic; most low mass galaxies will undergo star formation in the current time interval, with a range of star formation rates, but high mass galaxies either undergo a starburst or do not depending on their merger histories. When the effect of all transients is considered (figure  \ref{fig:irrmass}, right), the bias towards low irradiated fractions is suppressed. A relatively low rate of star formation is sufficient to trigger GRB activity, and each GRB has a large effect, resulting in a near-uniform distribution of irradiated fraction that is largely independent of stellar mass.

\begin{figure}
  \includegraphics[width=\columnwidth]{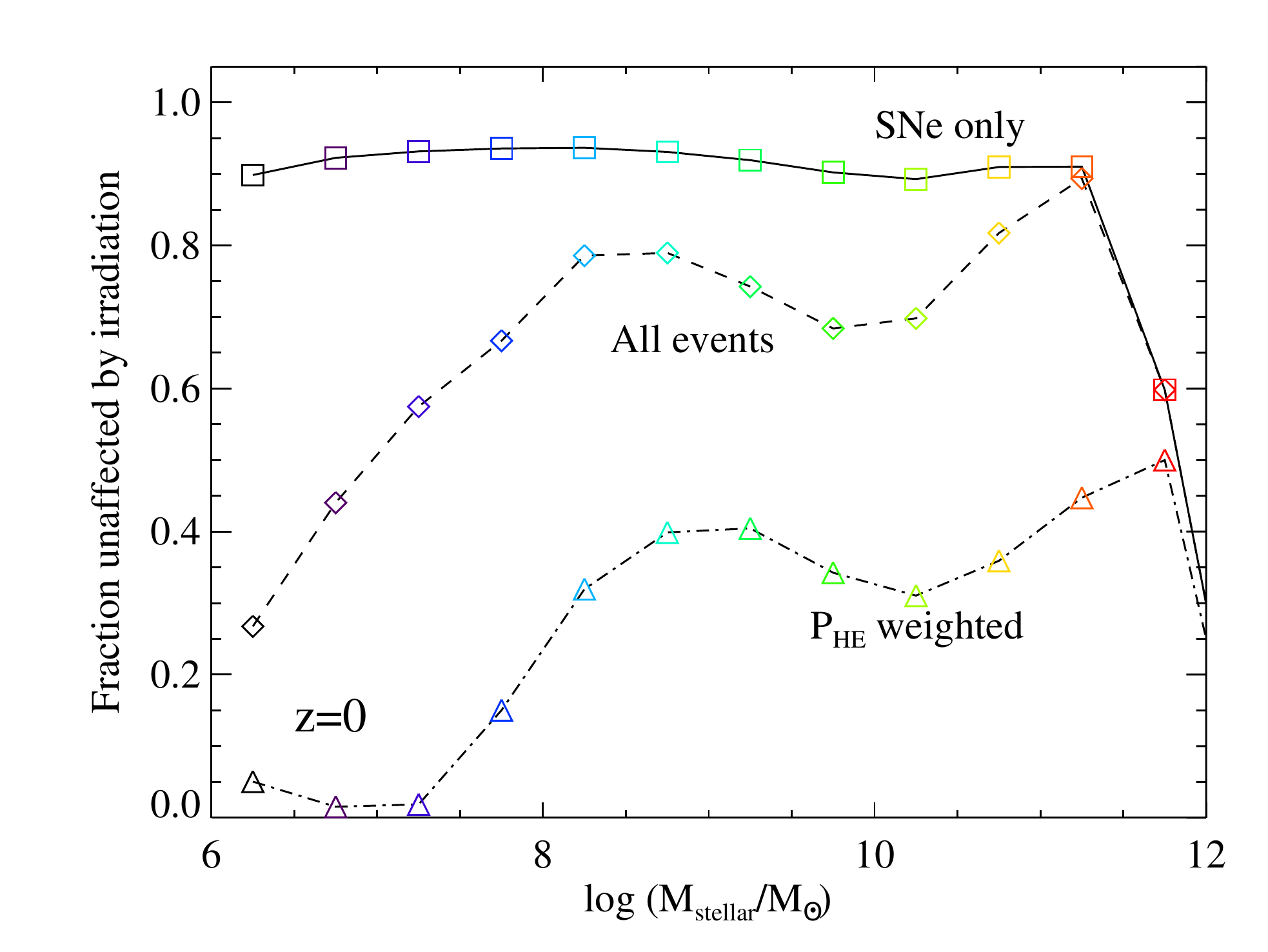} 
  \caption{The mean fraction that escapes irradiation in the current time interval as a function of stellar mass. We show the results when only SN irradiation is considered, and the effect of all transient events. We also show the mass fraction likely to both escape irradiation and host terrestrial planets (i.e. the P$_\mathrm{HE}$-weighted mass fraction)). This both reduces the overall fraction of stellar mass that may be consiered habitable and suppresses low mass galaxies.}\label{fig:irrmass2}
\end{figure}

If we consider the mean habitability fraction at each mass (figure \ref{fig:irrmass2}), the effect of the metallicity weighting $P_\mathrm{HE}$ becomes clearer. Given the strong mass-metallicity relation built into the Millenium Simulation, smaller galaxies will have typically lower metallicities and thus a lower probability of hosting terrestrial planets, so only a very small fraction of stars in low mass galaxies will both be capable of hosting terrestrial planets and escape irradiation. However at higher masses the mass-metallicity relation is shallower and thus the fraction of stars hosting potentially habitable terrestrial planets is a near-constant fraction of half the fraction escaping irradiation, consistent with $P_\mathrm{HE}$ at the mean metallicity at $z\sim0$. 

Galaxies at intermediate masses, log($M_\ast$/M$_\odot$)=9.5-10.5, show a deficit in potentially habitable mass fraction at the current time. This does not arise primarily from the effects of supernovae, but is rather dominated by the volumes irradiated by gamma ray bursts at these relatively late times. Interestingly the potential habitability actually peaks at a stellar mass log($M_\ast$/M$_\odot$)$\sim$11-12, comparable to or above that of the Milky Way and Andromeda.  As we discuss in the next section, this may well reflect the relatively early formation epoch of such massive galaxies and their low current star formation rates.

\begin{figure}
  \includegraphics[width=\columnwidth]{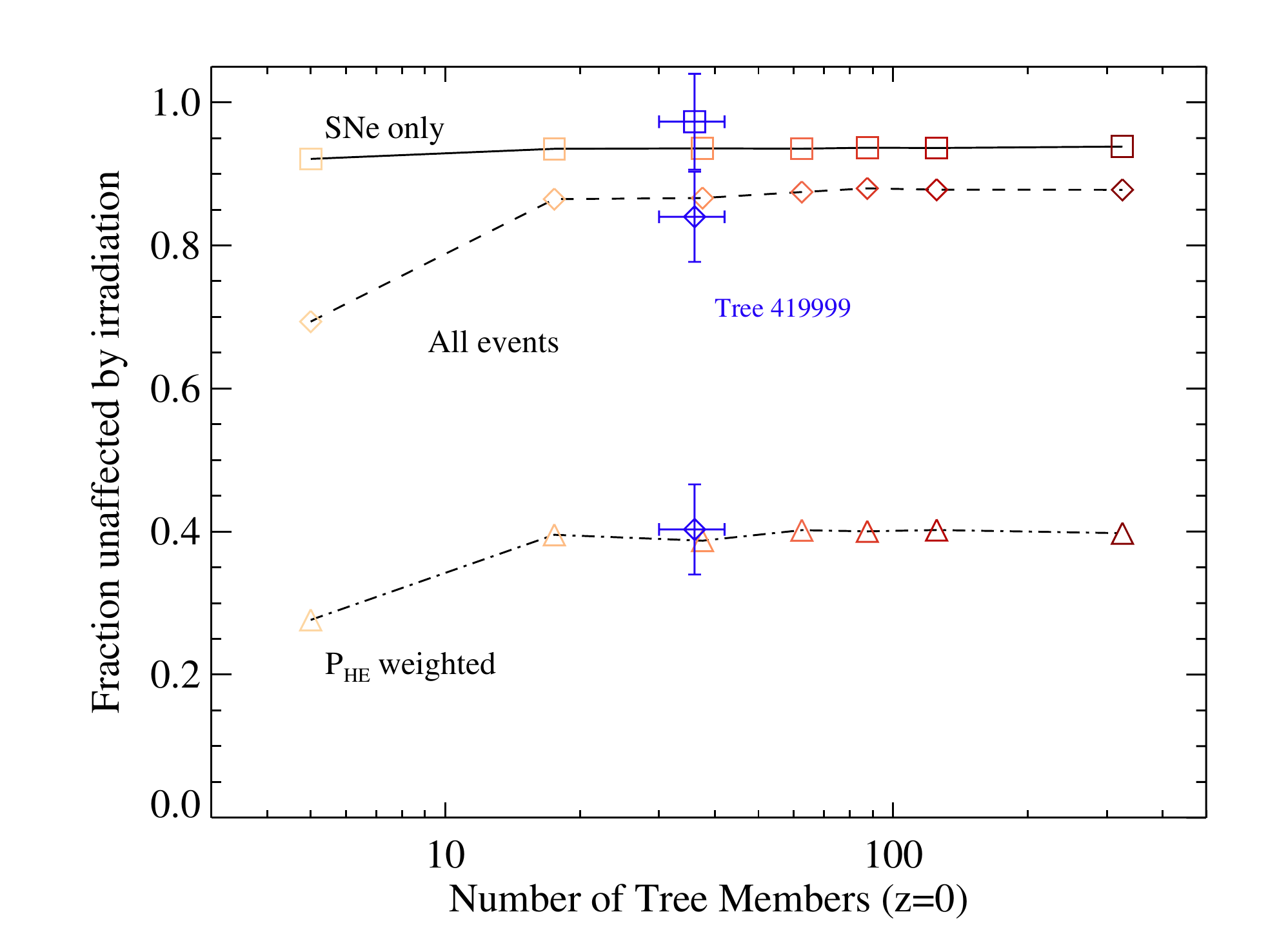} 
  \caption{The mean fraction that escapes irradiation in the current time interval as a function of richness of local environment, measured as the number of members of the merger tree at $z=0$. As before we show results when only SN irradiation is considered, and the effect of all events, as well as the fraction likely to host habitable planets while also escaping irradiation. The individual case study of merger tree ID 419999, discussed in section~\ref{sec:casestudy} is shown with blue points.}\label{fig:irrden}
\end{figure}

As figure \ref{fig:irrden} demonstrates, the richness of the local environment (in this case measured by the number of merger tree members identified in the simulation at $z=0$) has only a weak effect on the potentially habitable mass fraction of galaxies in the final simulation snapshot. In all but our lowest richness bin (of trees with 0-10 members at $z=0$), the mean potential habitability mass fraction is effectively constant. The far stronger effect on this bin of AGN and GRBs relative to supernovae suggest that this is a consequence of the typical galaxies in this bin being physically more compact, and so more affected by the large irradiation volumes of these energetic events. The least rich merger trees are the most abundant in the simulation at $z=0$ and they typically contain small galaxies, which are only just beginning to collapse into star forming systems at the end of the simulation, consistent with this interpretation. The compact sizes of these systems are therefore interpreted as reducing the average $H_\mathrm{rad}$ estimated in our model, which suggests that spatially extended galaxies, such as large disks and spirals, may be the most amenable regions in which to find life, regardless of larger scale group richness.

\section{Merger Tree Case Study}\label{sec:casestudy}

To explore the effect of galaxy merger histories on potential habitability in more detail, we consider a single merger tree as a case study.  Merger trees in the \citet{2007MNRAS.375....2D} analysis identify groups of galaxies through a friend-of-friends algorithm and tracks both the descendant of any given galaxy and its dominant progenitors. Here we consider as an example a single merger tree (Tree ID 419999) which has a total of 1925 individual members, of which 36 occupy the final snapshot (snap 63, $z=0$). These 36 sources range in stellar mass from $10^6$ to $10^{11}$\,M$_\odot$ and there are two members with masses exceeding $3\times10^{10}$\,M$_\odot$. Dwarf galaxies with masses $<10^6$\,M$_\odot$ are below the sensitivity of the simulation and are not considered (but may be considered to have low probable habitabilities given the low metallicities likely for such systems). The galaxies are distributed in a rough elipsoid with current dimensions $1\times1\times2$\,Mpc$^3$. As such this group represents a reasonable analogue to the Local Group of which our Milky Way and M31 are the dominant members.

\begin{figure*}
  \includegraphics[width=0.85\columnwidth]{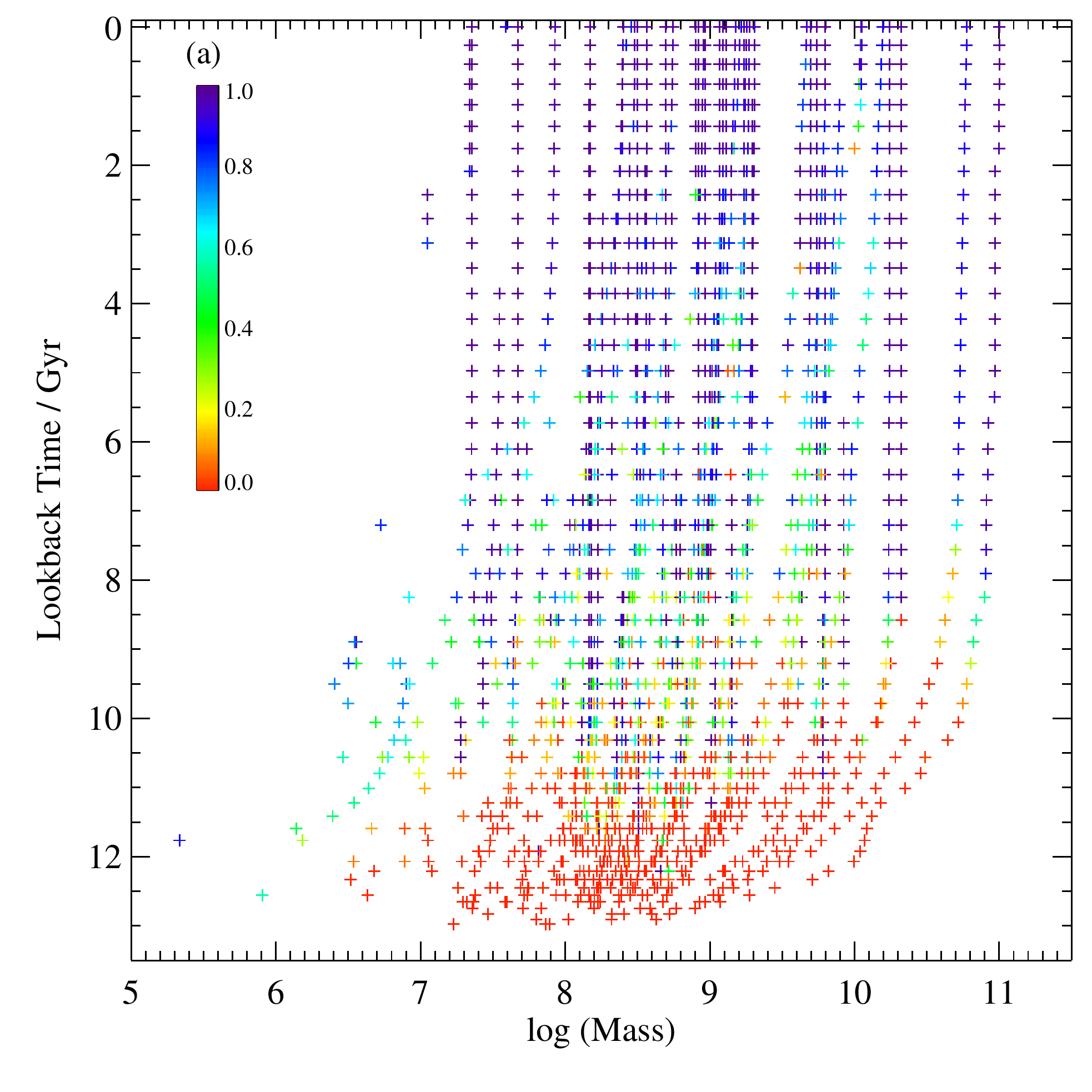} 
  \includegraphics[width=0.85\columnwidth]{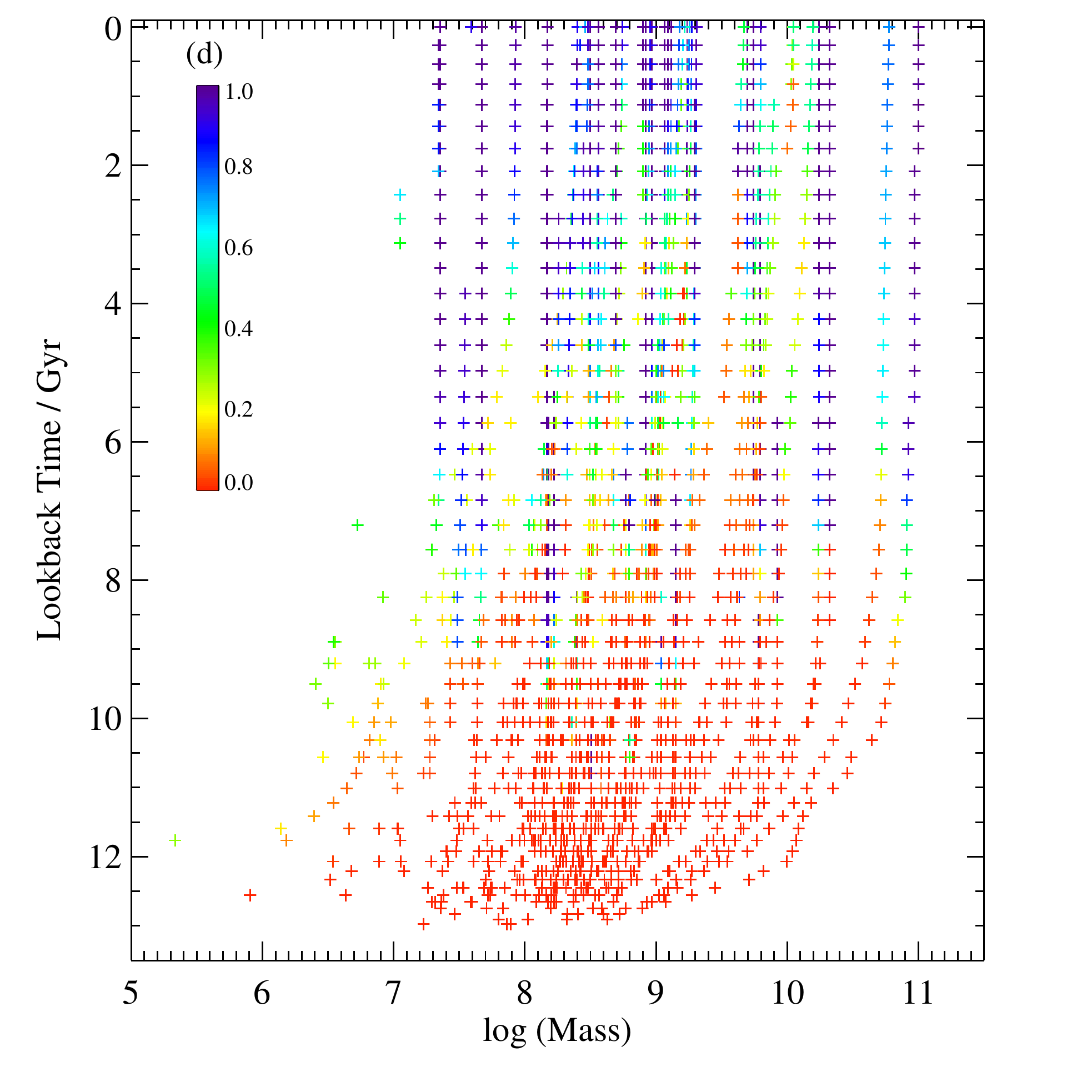} 
  
  \includegraphics[width=0.85\columnwidth]{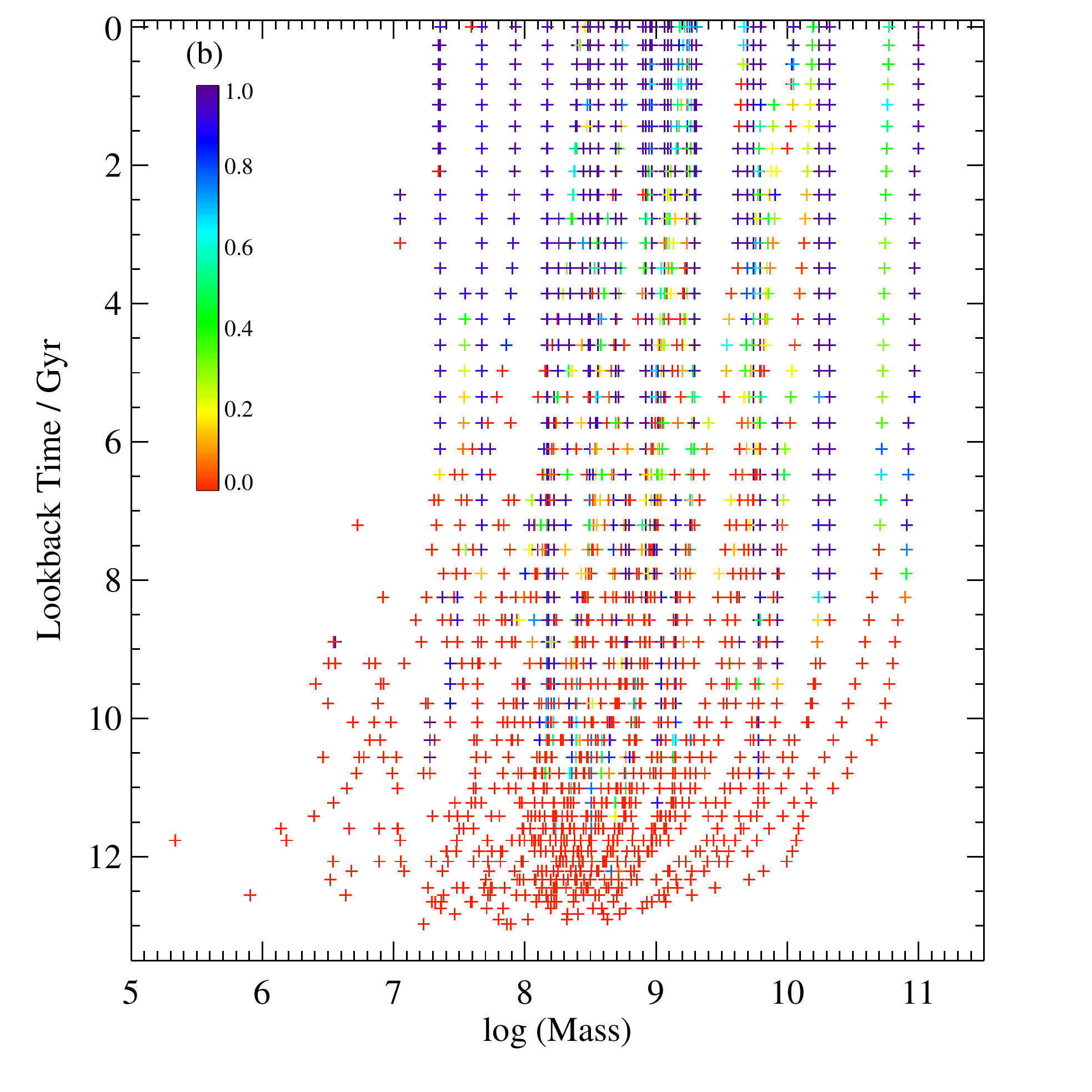} 
  \includegraphics[width=0.85\columnwidth]{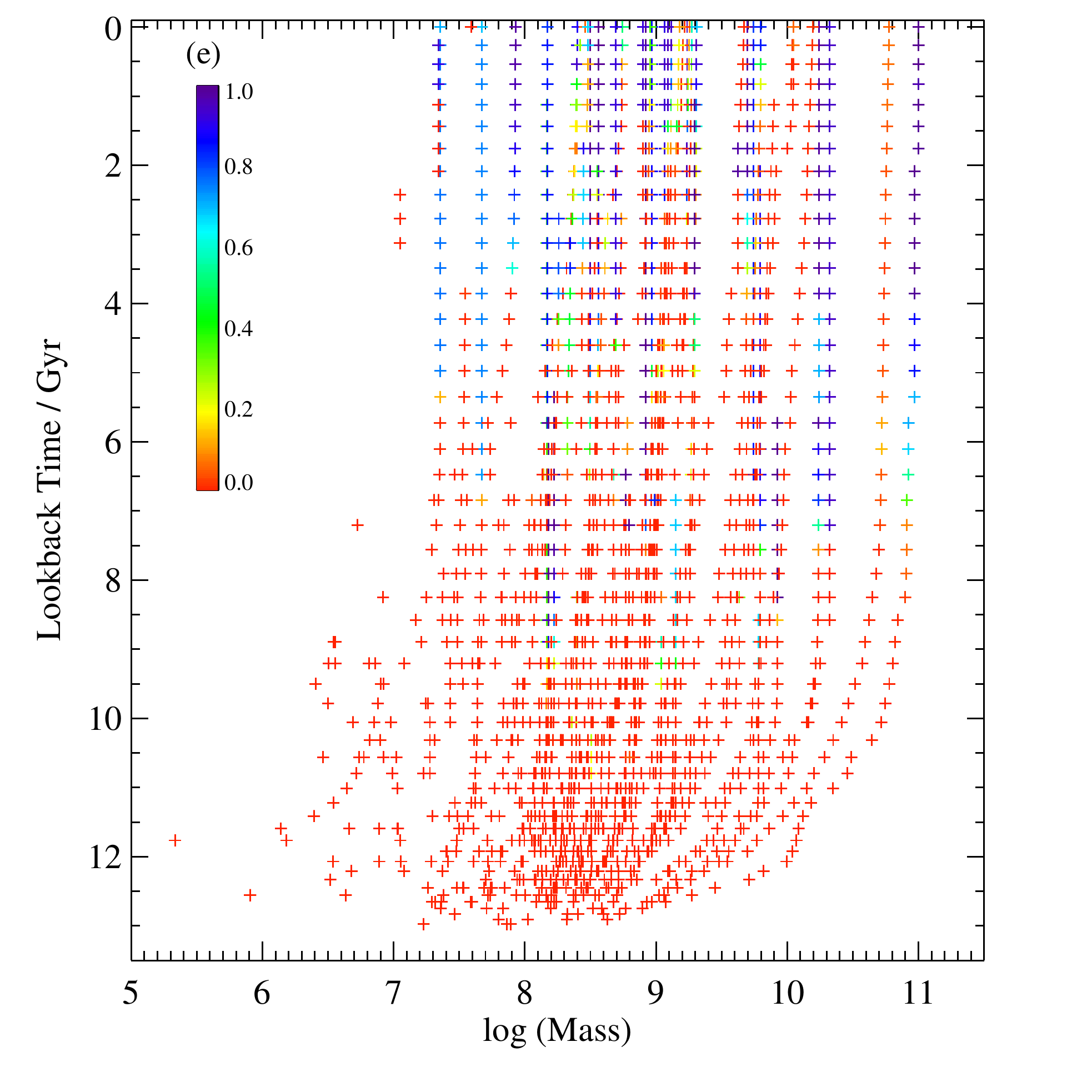} 
  
  \includegraphics[width=0.85\columnwidth]{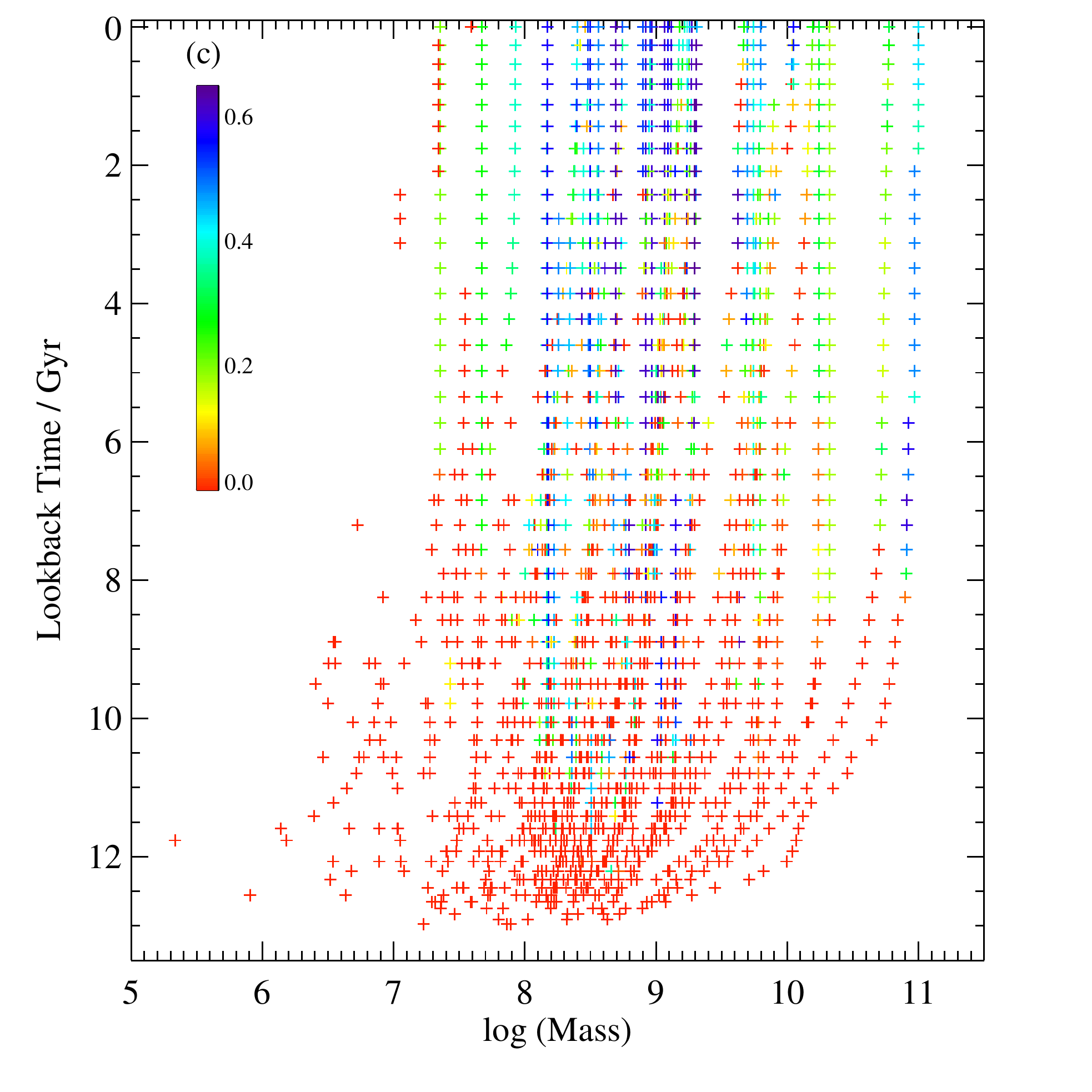} 
  \includegraphics[width=0.85\columnwidth]{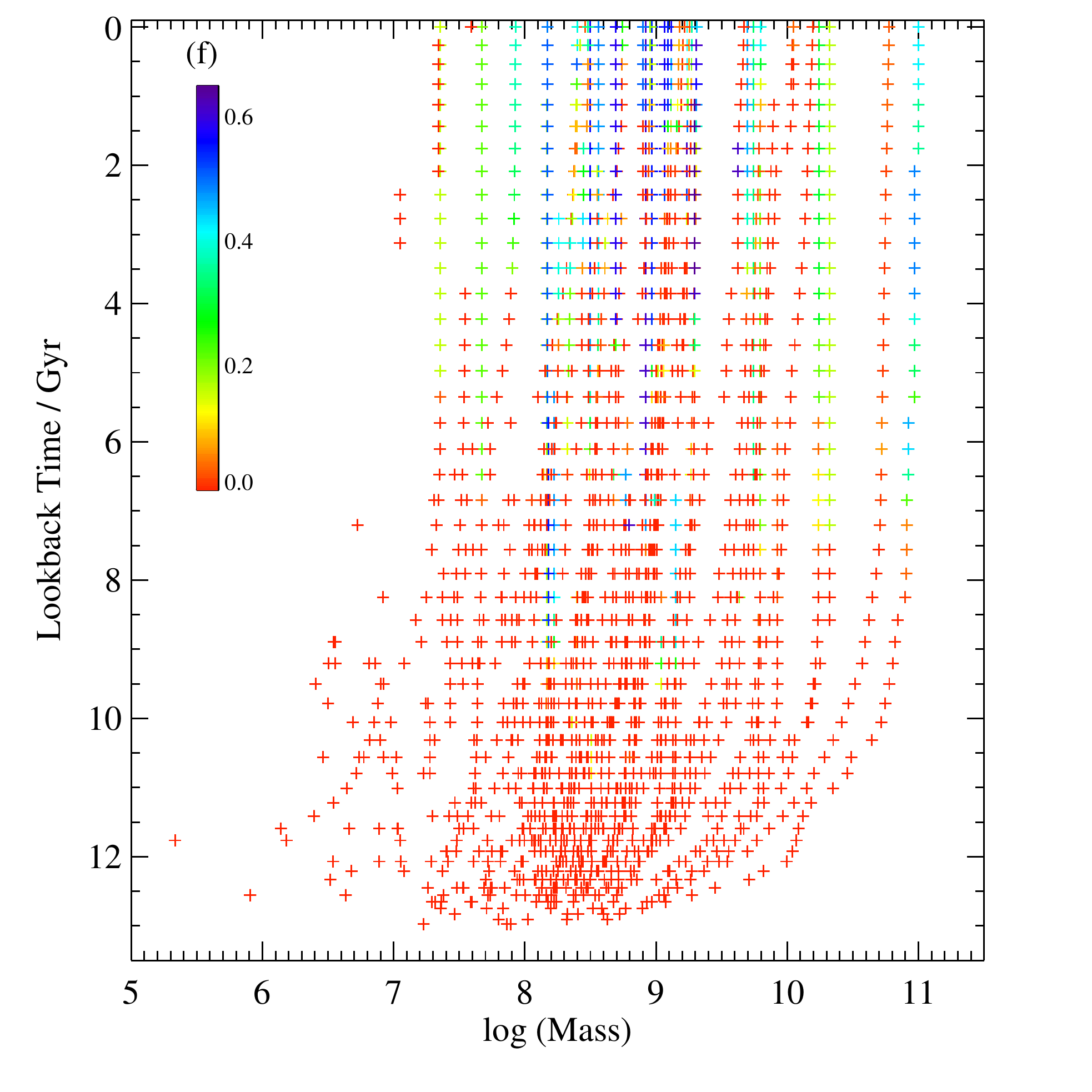} 
  \caption{The growth of mass with lookback time for galaxies in a selected, Local Group analogue, merger tree. Points are colour coded by a) the fraction of stellar mass which escapes irradiation by supernovae, b) the fraction of stellar mass which escapes irradiation by any transient source and c) the fraction of the stellar mass which may host unirradiated terrestrial planets, all in the time bin plotted. Panels d), e) and f) repeat this sequence, but giving the fraction in a rolling 1\,Gyr time interval preceding the snapshot in question. Note the change in colour-scale for panels c) and f) which weight irradiation-free mass fraction by terrestrial planet-hosting probability to obtain the habitable fracton $H_\mathrm{rad}$ at each time step.}
  \label{fig:tree1}
 \end{figure*}

In figure \ref{fig:tree1} we follow the mass evolution of the galaxies in this merger tree with lookback time, and colour code each galaxy by its potential habitability, $H_\mathrm{rad}$, at a given lookback time, based first on escaping irradiation within that time interval by considering supernovae (panel a) and second considering the cumulative impact of SNe, GRBs and AGN (panel b), before correcting the remaining, unirradiated fraction for the probability of hosting terrestrial planets, $P_\mathrm{HE}$ (panel c). This visualisation effectively captures both the global evolution of the population and the effects of mass build up and merger events on the irradiated stellar mass fraction. 

As expected from the analysis of volume-averaged quantities in the previous section, we find that the overall fraction that remains potentially habitable if only supernovae in the current time interval are considered (figure \ref{fig:tree1}a) is very high ($>90$ per cent) for the vast majority of galaxies. It is interesting to note that this remains true through the bulk of cosmic history for this merger tree and that irradiation-free fraction in any given galaxy rarely drops below 40 per cent within the last 8\,Gyr from the present time. If the additional impact of AGN and GRBs is considered however (figure \ref{fig:tree1}b), the overall picture becomes far less clear. Most individual galaxies show stochastic intervals of star formation or accretion activity which leads to individual snap intervals where the galaxy's irradiation-free mass fraction drops below 10 per cent. This occurs at all galaxy masses and at all cosmic times. Accounting for the fraction of irradiation-free stellar mass likely to host terrestrial planets  (figure \ref{fig:tree1}c) does little to change the overall pattern of habitability evolution but acts as a systematic reduction on habitable mass fraction.  Several galaxies show sustained high habitable fractions which are abruptly truncated with near-total irradiation, although some galaxies can escape significant irradiation for extended time periods.

The most massive galaxy in the simulated group, for example, experiences only minor fluctuations in its potentially habitable mass fraction in the last 8 Gyr of the simulation. However the second most massive galaxy experiences more  variation in habitability from one time interval to the next. Given its significant contribution to the total stellar mass in the group (this galaxy alone contributes 21 per cent of the mass), the greater variability of this source can have a substantial effect on the group's average properties.

 We explore the effects of timescale in the right-hand panels of figure \ref{fig:tree1} (panels d, e and f). Here instead of considering $H_\mathrm{rad}$ based on irradiation within a given snap interval, we show instead the fraction of mass which has not been irradiated within the last $t_\mathrm{h}=1$\,Gyr, in order to allow for the possible timescales required for the evolution of complex life \citep[rather than potentially rapidly-arising single-celled organisms,][]{2017Natur.543...60D}. As might be expected, the requirement that a star has not been irradiated within $t_\mathrm{h}$ reduces the fraction of the stellar mass in the group deemed plausibly habitable, even before correcting for the planet-hosting metallicity bias. The vast majority of progenitor galaxies for this group were essentially uninhabitable at lookback times greater than 6\,Gyr. Supernovae alone are now sufficient to drive the $z=0$ irradiation-free fraction of five of the 36 galaxies below 60 per cent, while the addition of GRB and AGN add an additional eight galaxies to this category. Weighting by the probability of hosting terrestrial planets suggests that these 13/36 galaxies have overall habitable mass fractions of under 20 per cent, with 10/36 below 10 percent.
 These include the single massive galaxy discussed above, which in this scenario is most analogous to the Milky Way and has a current time habitable fraction of 0.03 when the effects of all sources and the metallicity bias of terrestrial planet hosts are considered.

Despite this generally pessimistic view of potential habitability over an extended period in this group, it is worth noting that there are exceptions.  The most massive galaxy in the group formed the majority of its stellar mass at $z>2$ and has had a $H_\mathrm{rad}$ exceeding 40 per cent for 6\,Gyr. There are examples of galaxies at every mass scale which have retained a similarly high habitability fraction for a similar period, and in some cases for as long as 10\,Gyr. These are systems which have relatively simple merger trees after their early formation, and have not undergone substantial starbursts or AGN accretion episodes since $z\sim1$. In fact the main effect of requiring a substantial epoch free of irradiation is to reduce the continuous distribution of habitable fractions to a strongly bimodal distribution: either a galaxy has a $H_\mathrm{rad}>30$ per cent at $z=0$ or $H_\mathrm{rad}<5$ per cent, with very few sources lying in between.

\section{Uncertainties and Implications}\label{sec:discussion}

This is the first analysis to use the full merger tree, galaxy size and star formation history of a representative cosmic volume, and detailed theoretical predictions for transient event rates as a function of metallicity, to quantify the potentially habitable fraction of star forming galaxies as a function of mass and large scale environment. This represents a significant improvement on analyses which apply volume-averaged assumptions and scaling factors.  However there are areas where it has nonetheless been necessary for us to make assumptions, and these should be examined.

Perhaps the most significant assumption we make is to model each galaxy as a spherical distribution of stars with an exponential density profile. In fact, it is likely that galaxy morphology is far more complex throughout cosmic history. In the distant Universe, galaxies are typically clumpy and asymmetric, showing evidence for mergers or other interactions. By about $z\sim2$ (a lookback time of 10\,Gyr from the present time), massive disks were beginning to form, but a diversity of stellar distributions remain. In the local Universe, the familiar giant ellipticals and disks are accompanied by a range of irregular, spheroidal and dwarf galaxies. Any of these systems can have extensive substructure, including spiral arms, nuclear rings, bulges, and central cusps or cores. Incorporating this variation in morphology is not possible given the \citet{2007MNRAS.375....2D} semi-analytic model on which our analysis is built, and we are forced to adopt the simple parameterisation. We note that this is slightly shallower than the $r^{-2}$ density profile suggested by some authors, but scaling by each galaxy's virial radius has a far stronger effect than the precise distribution we adopt.

Our second major assumption lies in our adopted effective radii for astrophysical transients: $r_\mathrm{SN}$=8\,pc, $r_\mathrm{GRB}$=1\,kpc (with opening angle 10\degr) and $r_\mathrm{AGN}$=100\,pc. 

Of these, the supernova radius is the best established in the literature, based on detailed studies of an atmospheric response to ozone depletion \citep[][and later work]{1995ApJ...444L..53T} for a typical event. Nonetheless it remains unclear whether a more distant supernova could plausibly have biological implications for Earth-like planets and there is evidence on Earth itself of supernova impacts on statigraphic data. \citet{1999NewA....4..419F} have proposed an association between cosmic ray bombardment from a supernova event at 30\,pc and a mass extinction event 5\,Myr ago, based on measurements of $^{60}Fe$ in deep ocean crust. Similarly \citet{2006AdSpR..38.1307D} have reported detectable spikes in the nitrate concentration of polar ice cores, due to ionizing radiation flux associated with historical supernovae known to have occured at kiloparsecc distances. While these events clearly did not destroy all complex life on Earth (as our own existence testifies), the integrated impact of mid-distance (8-1000\,pc) events on climate, genetic stability and reproductive success could potentially be substantial in galaxies with high star formation rates where such events are common. Indeed, even in the Milky Way (with a relatively low star formation rate), \citet{2002ApJ...566..723S} have estimated that $\sim100-500$ events may affect biology on Earth per Gyr. Further study is required into the potential effects of irradiation on organic life as a function of distance and a more complex, future analysis may wish to apply a tapering impact distance function, rather than a simple cutoff.

Our analysis also assumes that each supernova irradiates a near-identical volume. There are reasons to believe this may not be true. SNe have a wide range of luminosity which would be reflected in their irradiation volumes. Perhaps the most extreme instance of this would be superluminous supernovae (SLSNe). There is probably only one SLSN for every $\sim$1000 core collapse events locally \citep[e.g.][]{2013MNRAS.431..912Q,2017MNRAS.464.3568P}, and they are likely subject to a strong metallicity bias, but are 50-100 times brighter, and so illuminate a volume that is about 1000 times as large. As they also produce more UV radiation (due to relatively little metal line blanketing), the habitability consequences could be quite significant. A counter influence may be provided by the local environment of supernovae. The early stages of star formation are usually heavily enshrouded by dust (e.g. graphite or silicate grains) in its immediate vicinity \citep[see e.g.][]{2017MNRAS.470.4453R}. It is possible that this provides some degree of protective effect from SNe that occur on timescales of $\lesssim10$\,Myr after the onset of star formation.

We also note that there are a range of smaller events which have the potential to irradiate Earth or Earth-like planets. These include the explosive energy releases accompanying intermittant accretion onto stellar-mass compact objects (cataclysmic variables and X-ray binaries) and flares from magnetars \citep[including the subclass of soft gamma repeaters which display occasional giant flares, e.g.][]{1999ApJ...527L..47W}. These can release hard X-rays which are potentially harmful to life if occuring sufficiently close to a terrestrial planet. Magnetar flare events are relatively rare \citep[e.g][]{2015MNRAS.447.1028S}, since the evolutionary pathways leading to magnetar creation and flaring are much weaker than those of massive stars leading to core collapse supernovae. In old stellar populations, however, cataclysmic variables (accretion onto white dwarfs) or type Ia supernovae (the thermonuclear detonation of a white dwarf) may be a dominant source of hard radiation, and this is not fully accounted for in our study.
We do not consider the full luminosity range of cosmic explosions, or dust enshrouding, in this paper but defer it to future work. 

The radius of influence of a GRB is also difficult to model. GRBs have a large range of peak energies and durations, and likely also in jet opening angle \citep[see][and references therein]{2003ApJ...599..408B,2007A&A...471..585B}. Other authors considering habitability have used an opening angle up to 20\degr, and a radius as high as 5\,kpc. To some extent the uncertainties in these are degenerate. Our selected values lie somewhere in the centre of the range for these systems - again assuming an Earth-like atmosphere.

Finally the AGN radius is still more difficult to constrain. \citet{2016arXiv160609224D} treated the region of influence of an AGN as identical to that of a supernova which is reasonable if peak energies are considered, but not when considering the sustained irradiation from these sources. By contrast, \citep{2016A&A...592A..96G} used an estimated for $r_\mathrm{AGN}$ which scales with stellar mass as $\log(r_\mathrm{AGN}$/kpc$)=-6.1 + 0.7 \times \log(M_\ast$/M$_\odot)$. This varies from 60\,pc at $M_\ast=10^7$\,M$_\odot$ to 8\,kpc at $M_\ast=10^{10}$\,M$_\odot$ and does not account in any way for jetted emission from the accretion disk or the protective effects that might be offered by an AGN's dusty torus \citep[see][for further discussion]{2011AsBio..11..343M}. These suggest that an opening angle adjustment, such as that used for GRBs might be appropriate, but an analysis of the biological effects from such emission is beyond the scope of this work.

These volumes have a large effect on the irradiated mass fraction predicted in our models and the relative contribution of different components (see appendix for further discussion), but adjusting them acts primarily as a multiplier on $H_\mathrm{rad}$. In other words, the final potentially habitable stellar mass fraction of any given galaxy may be slightly higher or lower than we predict, but the general trends seen in the analysis above are unlikely to change significantly. 

The timescale for habitability ($t_\mathrm{h}$=1\,Gyr in our analysis), presents another consideration. \citet{2017Natur.543...60D} have used evidence from hydrothermal vents to suggest that life was present on Earth 4.28\,Gyr ago, only a hundred Myr after Earth's first liquid water formed. \citet{1999JBIS...52...19A} has suggested that a timescale of this order might be appropriate for the development of intelligent life. However it took $\sim$2.5\,Gyr more for life on Earth to make the jump to multi-celled organisms and a further Gyr before intelligent life arose. Extrapolating from a sample of one is always problematic, but it seems probable that a minimum interval of 1\,Gyr is required to develop a reasonably complex organism that might be straightforwardly recognisable as life, or which might generate exoplanet biosignatures. A Gyr timescale is also relevant for the irradiation arguments in this study. Assuming an 8\,pc irradiation radius for supernovae, \citet{2011AsBio..11..343M} argue that Earth may experience 1-2 extinction level radiative events per Gyr, consistent with identification of the Ordovician extinction event ($\sim$440\,Myr ago) as the latest of these. However this interpretation is not universally accepted. 

Finally, as noted in section \ref{sec:habZ}, the true probability weighting for hosting terrestrial planets as a function of metallicity remains uncertain, although the effects of this weighting are reflected more in the overall scaling of the habitable fraction than in its evolution with cosmic time. All of the preceeding presupposes that life requires a terrestrial world (or exomoon) with an atmosphere much like our own for its survival - something that is challenging to justify given the variation in atmospheric conditions over the history of life on Earth, and does not take into account potential variation elsewhere.

Accepting these caveats, what can we conclude about the evolution of cosmic habitability?  Early work considering the effects of GRBs by \citet{1999JBIS...52...19A} suggested that a burst capable of destroying intelligent life might occur as often as every 100\,Myr. Also considering the effects of GRBs, \citet{2014PhRvL.113w1102P} estimated that more than 95 per cent of the inner 4\,kpc of the Milky Way has been irradiated by GRBs in the last 5\,Gyr, with only the outskirts of the galaxy remaining habitable. This is broadly consistent with our finding that galaxies comparable to the Milky Way can have very low potential habitability fractions when an extended epoch is taken into account.
Perhaps more relevant to the cosmic timescales considered in the Millenium Simulation merger trees is the assertion of  \citet{2014PhRvL.113w1102P} that the Universe was essentially uninhabitable at $z>0.5$ (lookback times $>$5\,Gyr). \citet{2016arXiv160609224D} also suggested that the habitability has evolved rapidly since Earth formed, with potentially a 40-fold increase in habitable fraction in the last 4\,Gyr. \citet{2016A&A...592A..96G} disagreed with both these studies, suggesting that the average habitability of the Universe hasn't changed significantly in 8\,Gyr. As figure \ref{fig:irrtime} demonstrated, we find an intermediate result. While we agree with previous studies that the early Universe was extremely hostile to life, we estimate that at a lookback time of 4-5\,Gyr from the present time only 50 per cent of the stellar mass in a volume-averaged sample is likely to have been irradiated within a $\sim200$\,Myr interval. Extending this interval to 1\,Gyr increases the mean irradiated fraction, but, as shown in figure \ref{fig:tree1} and discussed in section \ref{sec:casestudy}, there is substantial variation between the irradiated fractions of galaxies at any given mass, in addition to scatter in the metallicity-dependent $P_\mathrm{HE}$.  We can even identify galaxies which experienced an extended habitability period of $\sim2$\,Gyr or longer early in the Universe but which have since undergone a near-certain extinction event - an example is visible in the merger tree in figure \ref{fig:tree1} at log($M_\ast$/M$_\odot$)$\sim8$.

So where is the best place for life like our own to survive in the Universe? Our analysis of both a volume-averaged sample and a case study merger tree suggest that, when the complexities of star formation history and metallicity-dependent transient rate are taken into account, there is not a simple answer to this question.  The low redshift Universe is clearly less subject to catastrophic irradiation events than the Universe at early times, with a far higher fraction of the stellar mass in galaxies remaining clear. It appears however that neither a galaxy's stellar  mass (figure \ref{fig:irrmass2}) nor its large scale environment (as measured through group membership, figure \ref{fig:irrden}) is a strong factor in determining its potentially habitable mass fraction. While massive ($M_\ast\sim10^{11}$\,M$_\odot$) galaxies have the highest average habitable mass fraction at the present day, we caution that scaling the AGN irradiation radius by black hole mass (i.e. accounting for Eddington luminosity dependence) would affect these sources most strongly, and this result cannot be considered robust. 

A result we do consider both robust and interesting is the very clear differences in the habitability history of galaxies of the same mass demonstrated by figure \ref{fig:tree1}. This is introduced by a careful consideration of the galaxies' merger trees and so accounts for both star formation induced by mergers between galaxies with different mass ratios, and also for the metallicity evolution of the galaxies. Given the very strong dependence of GRB rates on metallicity (as shown in figure \ref{fig:transient}), this is crucial to securing an accurate overview of the cosmic evolution of galactic habitability.

\section{Conclusions}\label{sec:conc}

We summarise our main conclusions as follows:
\begin{enumerate}
\item We have combined the merger trees and galaxy properties arising from the semi-analytic model of \citet{2007MNRAS.375....2D} and N-body simulation of \citet{2005Natur.435..629S} with metallicity-dependent rates for astrophysical transients derived with the Binary Population and Spectral Synthesis model code of \citet{E17}.

\item Using these models, we have considered the evolution in galactic habitability (as measured by the mass fraction of each galaxy that escapes irradiation from energetic transients and the fraction of the remaining mass capable of hosting terrestrial planets) in a representative local volume in a specified time interval. We find that 18 per cent of stars are irradiated within the simulation interval ($\sim$260\,Myr) at $z\sim0$. This fraction rises to 50 per cent at $z=0.5$ and exceeds $95$ per cent at $z>2$.

\item We find that there is a large range of irradiated fraction at any given stellar mass, and that it is largely independent of the density of the galactic environment (as measured by the number of members of the galaxy merger tree at the present time).

\item The metallicity-dependent probability of hosting terrestrial planets suppresses habitability in the lowest mass (lowest metallicity) galaxies, but otherwise acts primarily as a scaling factor, reducing the habitable fraction to 40-60 percent of the irradiation-free fraction.

\item By considering the detailed merger tree of a representative Local Group analogue in the simulations, we explore the range of histories in galaxy habitability. We find that including AGN and GRBs as irradiating sources alongside supernovae has a significant effect on the habitability of individual galaxies, despite the relative rarity of these events. We also see that requiring an extended period free of irradiation reduces the number of potentially habitable galaxies at all masses.

\end{enumerate}

In all of the preceeding, we have made the assumption that life has similar biological requirements to that on our sole existing example, and also that it can only arise on terrestrial planets with similar magnetic environments and atmospheres to that of Earth. If we have learnt one thing from the recent boom in exoplanet discoveries it's that planets come in a vast range of sizes, shapes and environments \citep{2010ARA&A..48..631S,2015ARA&A..53..409W}. We should  expect life to be similarly diverse. Nonetheless, exploring the potential habitability of the Universe to lifeforms like our own helps to place the fact of life on Earth in context. The large range of habitable fractions in galaxies - even those similar in mass to the Milky Way - is testament to the vulnerability of planets to irradiation by astrophysical transients on cosmic timescales.

Finally, we note that we have demonstrated the application of galaxy merger trees to exploring the impacts of cosmic history on a galaxy's habitability. While the cosmological model we considered lacked the mass or spatial resolution to explore the variation of stellar structure, metallicity and its effects within a given galaxy, it has allowed us to obtain a more comprehensive overview of the average properties of a galaxy with known mass, metallicity and environment.  As computational power increases, N-body simulations, hydrodynamic simulations and semi-analytic models are improving our understanding of galaxy evolution, and are themselves being further constrained by new observational evidence \citep[e.g.][]{2014Natur.509..177V,2014MNRAS.444.1518V,2015MNRAS.446..521S,2017arXiv170703397S}. While  simulations designed to probe cosmological volumes are unlikely ever to resolve the parsec scales of individual star formation regions, they are producing an ever increasing number of model particles (or mass elements) per galaxy, and further work should allow more exploration of the diversity of properties in different subsets of the same galaxy halo, rather than between halos.  It is also becoming more common to model the evolution of individual isolated clusters and dark matter halos in more detailed, non-cosmological simulations \citep[e.g.][]{2017MNRAS.470.4186B}. Most of these are optimised to reproduce the properties of the Milky Way, Local Group or other nearby systems in detail \citep[e.g.]{2017IJAsB..16...60F,2017MNRAS.467.4383S}. While the habitability of the Milky Way and Andromeda in particular have been studied in detail using such models, and habitable zones estimated for them  \cite[e.g.][]{2016MNRAS.459.3512V,2017IJAsB..16...60F}, the increasing precision of galaxy evolution models could be used as a structure on which to base the estimation of habitability in a far wider variety of systems in the near future. This study has served as a pilot for such work and suggests that further investigation of habitability using cosmological simulations promises to be a productive area.

\section*{Acknowledgements}

We thank J.J. Eldridge for helpful discussions regarding the \bpass\ models and members of the Warwick interdisciplinary Centre for Exoplanets and Habitability for helping to inspire this work. We acknowledge undergraduate student summer research funding from the Royal Astronomical Society (ERS, MAL) and Warwick University (MAL, MJH). We also acknowledge PhD studentship funding from the UK Science and Technology Facilities Council (STFC) (SMLG, HJTC) and support from the Warwick Institute of Advanced Studies (GCB).

We thank the referee of this paper for helpful and thoughtful input on our manuscript. We acknowledge the substantial amount of work that went into the Millenium Simulation and the \citet{2007MNRAS.375....2D} semi-analytic models, and thank their authors for making their results public. The Millennium Simulation databases used in this paper and the web application providing online access to them were constructed as part of the activities of the German Astrophysical Virtual Observatory (GAVO).

\appendix

\section{Effects of Varying Irradiation Radii on Cosmic Habitability}\label{appendix}

\begin{figure*}
  \includegraphics[width=\columnwidth]{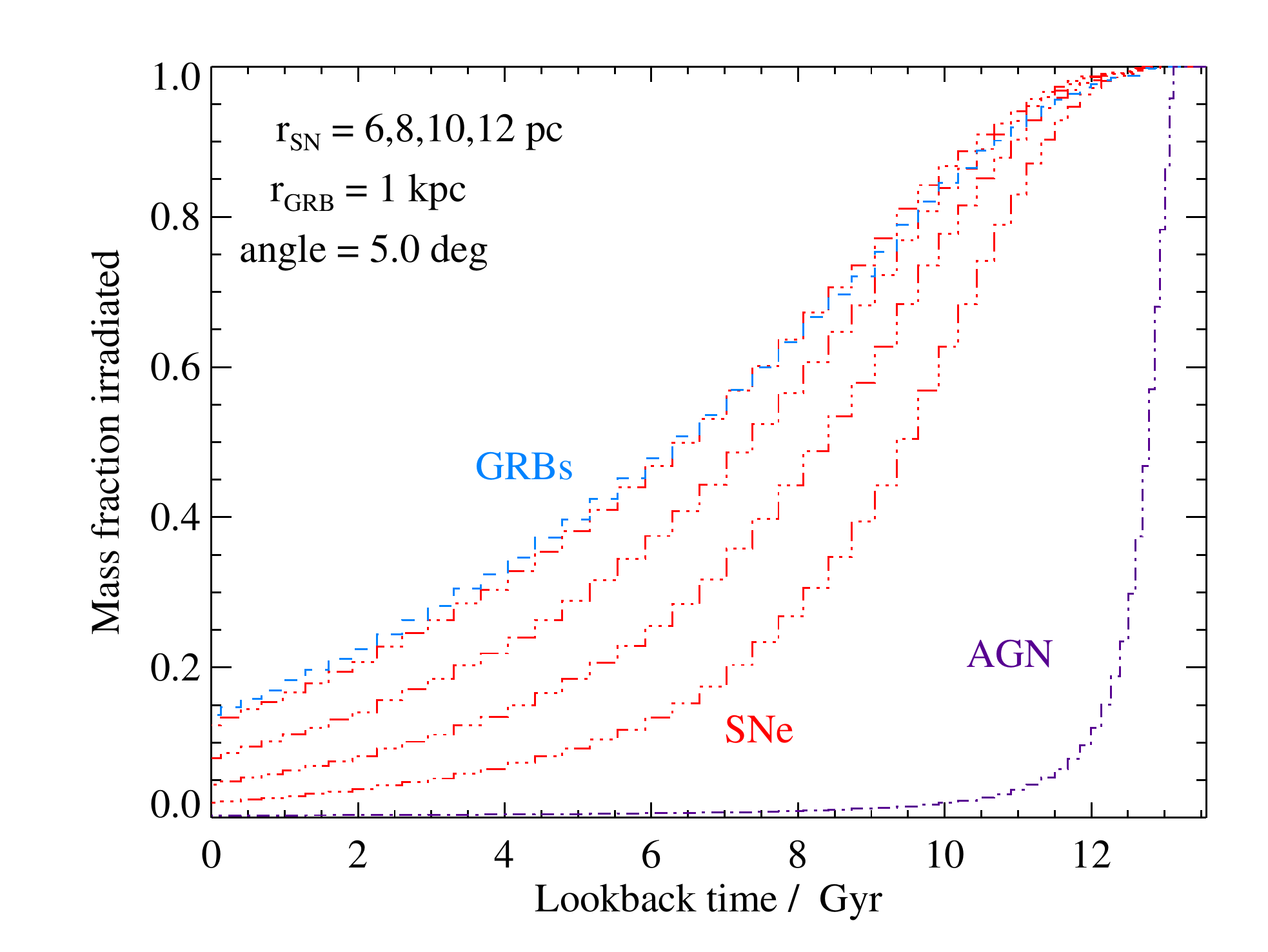} 
  \includegraphics[width=\columnwidth]{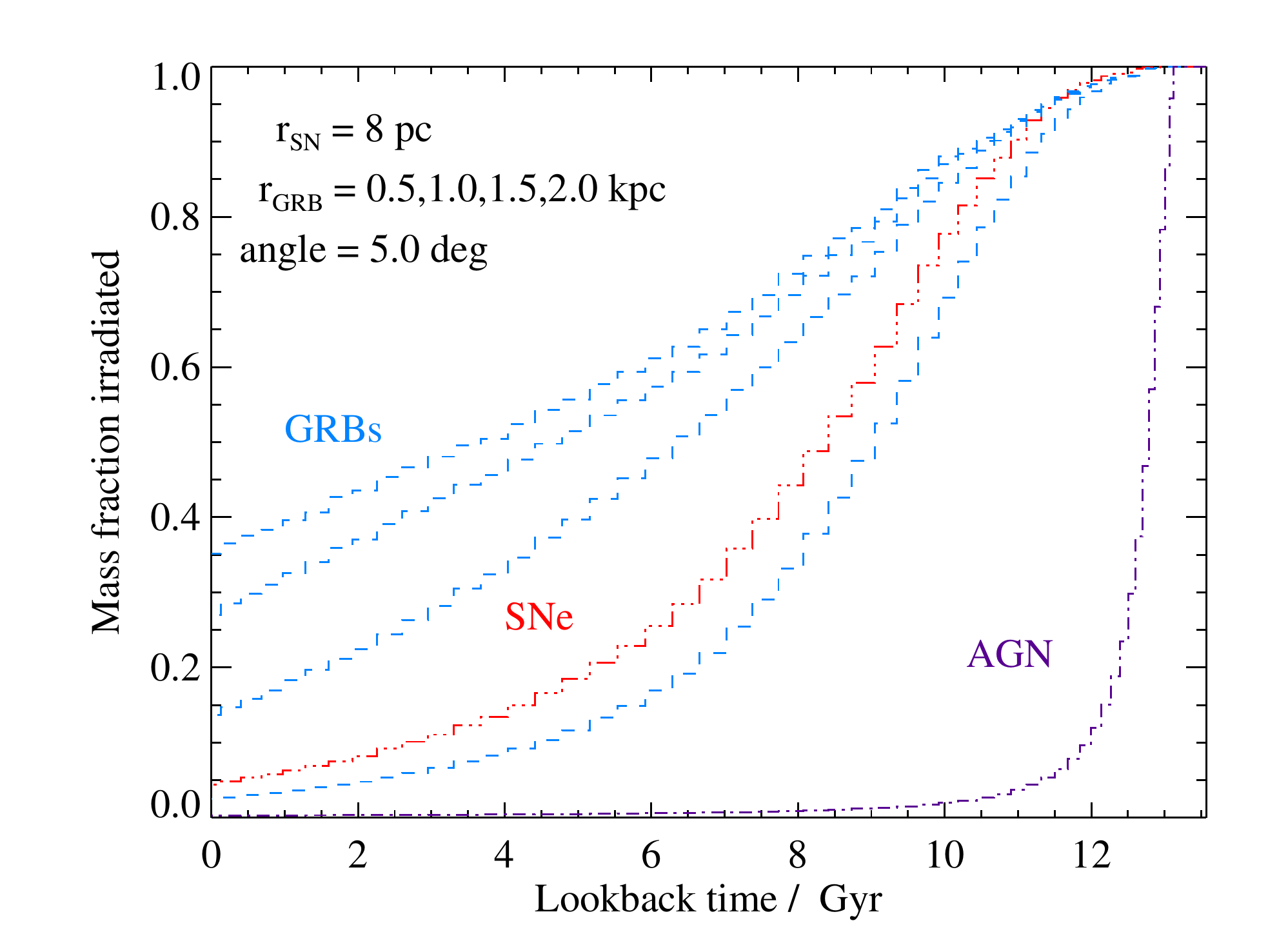} 
  \includegraphics[width=\columnwidth]{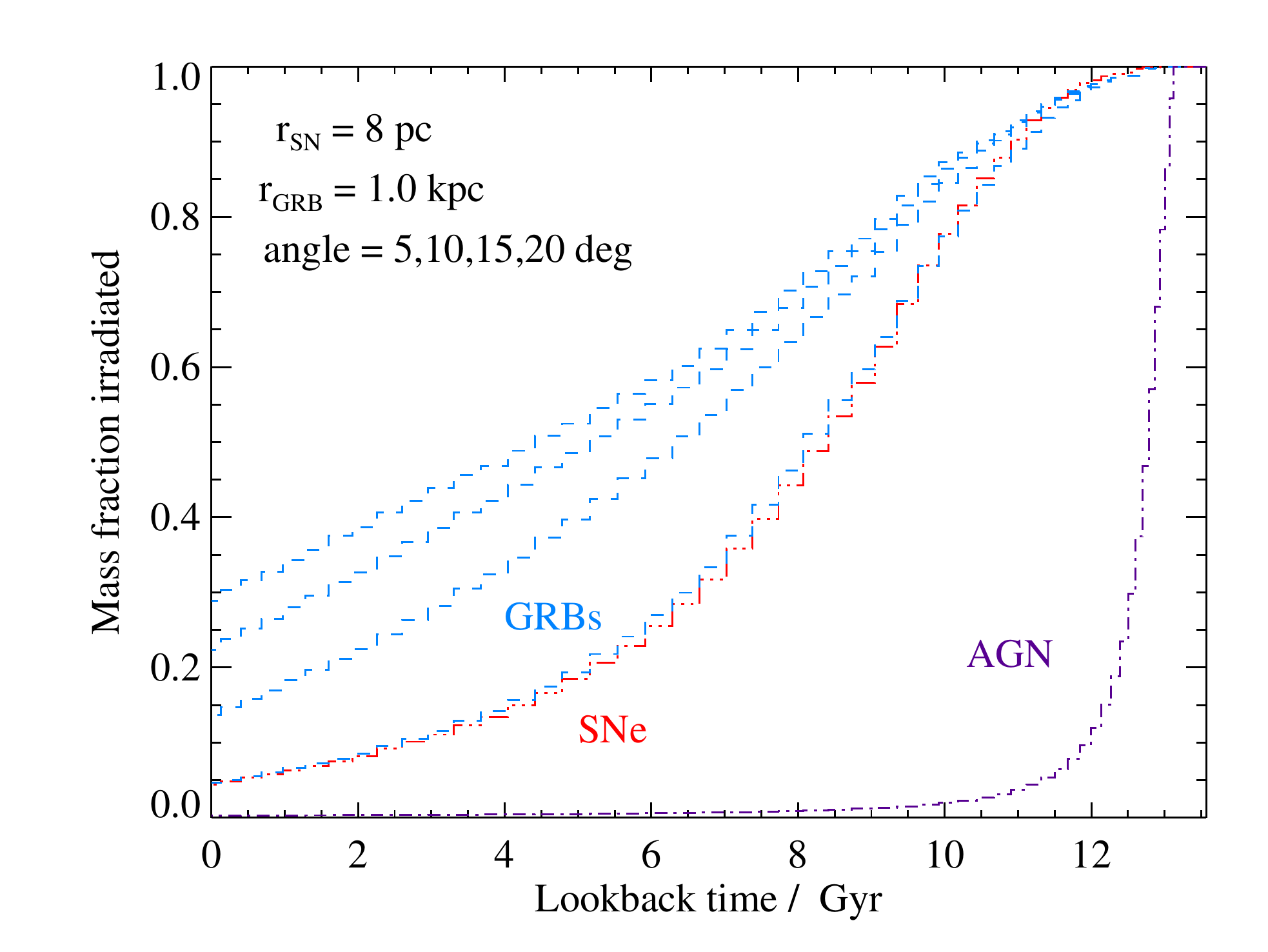} 
  \includegraphics[width=\columnwidth]{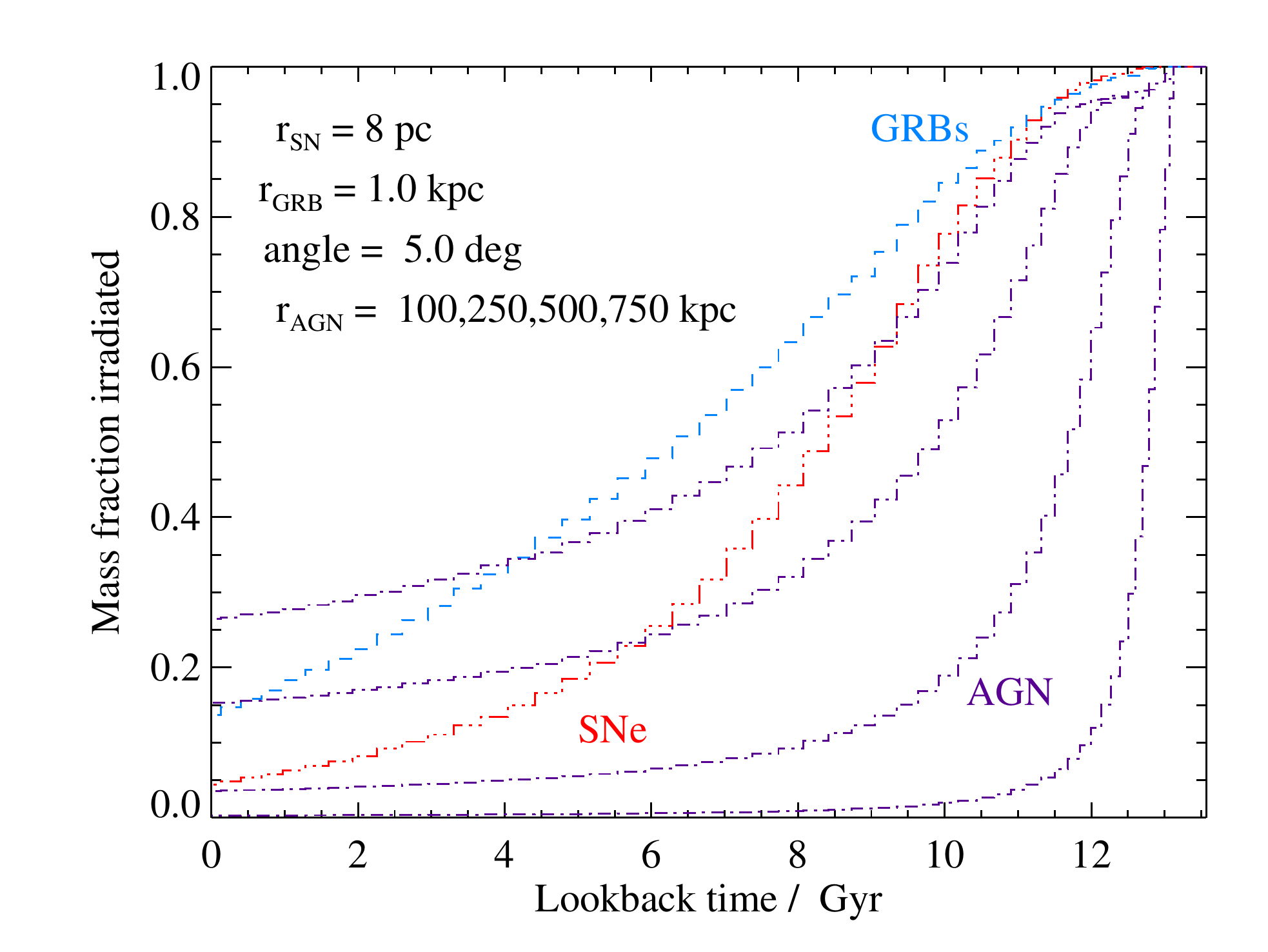} 
  \caption{The effect on volume-averaged irradiation fraction of changing the parameters governing the volumes irradiated by each events: (Top left) varying the radius of irradiation of supernovae showing $r_\mathrm{SN}$=6, 8, 10 and 12\,pc; (Top right) varying the GRB irradiation volume by changing the radius of influence, $r_\mathrm{GRB}$=0.5, 1.0, 1.5, 2.0\,kpc; (Bottom left) varying the GRB irradiation volume by changing the jet opening angle, $\theta$=5, 10, 15, 20\, degrees; (Bottom right) varying the region of influence of the central AGN, $r_\mathrm{AGN}$=100, 250, 500, 750\,kpc, assuming spherical irradiation.}\label{fig:radii}
 \end{figure*}

As discussed in section \ref{sec:discussion}, the detailed interpretation in this paper is based on an assumption regarding the volume irradiated by astrophysical transient events, and thus is subject to uncertainties associated with changing the volumes of influence. In figure \ref{fig:radii} we demonstrate the impact of changing these volumes on the volume-averaged mass fraction of stars irradiated in any given simulation snap interval.

We find that supernovae with a radius of influence $r_\mathrm{SN}$=12\,pc can match the affect of our assumed GRB model on the volume-averaged habitable mass fraction. Their small individual volumes are compensated for by their much higher rate relative to the more energetic explosions. We note that a major assumption in our analysis is that each SN irradiates a similar volume, as discussed in the main text. 

The volume of influence of GRBs depends on both the radial distance over which sources are affected and by the opening angle of jetted, relativistically beamed emission. Increasing both quantities increases the irradiated mass fraction, leading to GRBs becoming increasingly dominant at low redshifts and lookback times. However, these two parameters are likely to anticorrelate - a more tightly beamed GRB jet will be effective over a larger distance and vice versa. Our default choice lies towards the middle of the plausible range of GRB affects, but we caution that these vary over a factor of $\sim$6 at $z\sim0$.

The AGN irradiation volume is perhaps the least well constrained by previous studies. A simple galaxy-specific estimator might be based on the bolometric luminosity of its central supermassive black hole accreting at the Eddington limit. This will add 10 per cent to the bolometric luminosity of the Sun at Earth if a black hole of $10^6$, $10^7$, $10^8$, $10^9$\,M$_\odot$ were within $r=$2.8, 8.8, 28 and 88\,pc respectively (i.e. $r_\mathrm{AGN}/\mathrm{pc}=2.8\times10^{-3}\,\sqrt{M_\mathrm{BH}/\mathrm{M}_\odot}$). However it is far from clear what impact a 10 per cent boost in solar irradiation would have on life, particularly given that AGN will typically have a harder ionizing spectrum. Arguably a much lower irradiation boost would be sufficient to push a planet out of the habitable zone. Estimates used in the literature have varied between the same $\sim$10\,pc radius as a supernova \citep[e.g.][]{2015ApJ...810L...2D} to an $r_\mathrm{AGN}$ which scales with stellar mass as $\log(r_\mathrm{AGN}$/kpc$)=-6.1 + 0.7 \times \log(M_\ast$/M$_\odot)$  \citep{2016A&A...592A..96G}. This latter estimator varies from 60\,pc at $M_\ast=10^7$\,M$_\odot$ to 8\,kpc at $M_\ast=10^{10}$\,M$_\odot$. In both cases, there is an assumption of spherical irradiation, as mentioned in section \ref{sec:discussion} this may be inappropriate given the collimation of AGN jets and the shielding effect of a dusty circum-AGN torus. The volumes will be reduced by a factor of 0.01 for a 22\,degree opening angle \citep[typical of parsec scale AGN jets in the MOJAVE survey,][]{2017MNRAS.468.4992P}.

Given the above, our default choice of $r_\mathrm{AGN}$=100\,pc (spherical) was assumed as a reasonable estimate. In the figure we demonstrate the effect of increasing this. Given the centrally-concentrated stellar density distribution we assume, changing the volume affected by a central AGN can have a large effect, particularly at low redshift, where our default assumption irradiates a negligible mass fraction of large galaxies.

\bsp	
\label{lastpage}
\end{document}